\documentclass[aps,pre,preprint,show pacs]{revtex4}
\usepackage{graphicx}
\usepackage{dcolumn}
\usepackage{amsmath}
\usepackage{amssymb}
\usepackage{amsfonts}
\usepackage{rotating}
\usepackage[percent]{overpic}
\usepackage{bm}
\usepackage{color}
\newcommand{\be}{\begin{equation}}
\newcommand{\ee}{\end{equation}}
\newcommand{\ba}{\begin{eqnarray}}
\newcommand{\ea}{\end{eqnarray}}

\begin{document}
\title{The barrier to ice nucleation in monatomic water}
\author{Santi Prestipino}
\thanks{Email: {\tt sprestipino@unime.it}}
\affiliation{Dipartimento di Scienze Matematiche ed Informatiche, Scienze Fisiche e Scienze della Terra, Universit\`a degli Studi di Messina, Viale F. Stagno d'Alcontres 31, 98166 Messina, Italy}
\date{\today}
\begin{abstract}

Crystallization from a supercooled liquid initially proceeds via the formation of a small solid embryo (nucleus), which requires surmounting an activation barrier. This phenomenon is most easily studied by numerical simulation, using specialized biased-sampling techniques to overcome the limitations imposed by the rarity of nucleation events. Here, I focus on the barrier to homogeneous ice nucleation in supercooled water, as represented by the monatomic-water model, which in the bulk exhibits a complex interplay between different ice structures. I consider various protocols to identify solidlike particles on a computer, which perform well enough for the Lennard-Jones model, and compare their respective impact on the shape and height of the nucleation barrier. It turns out that the effect is stronger on the nucleus size than on the barrier height. As a by-product of the analysis, I determine the structure of the nucleation cluster, finding that the relative amount of ice phases in the cluster heavily depends on the method used for classifying solidlike particles. Moreover, the phase which is most favored during the earlier stages of crystallization may result, depending on the nucleation coordinate adopted, to be different from the stable polymorph. Therefore, the quality of a reaction coordinate cannot be assessed simply on the basis of the barrier height obtained. I explain how this outcome is possible, and why it just points out the shortcoming of collective variables appropriate to simple fluids in providing a robust method of particle classification for monatomic water.

\end{abstract}
\pacs{64.60.qe, 64.70.dg, 68.08.-p}
\maketitle

\section{Introduction}

Controlling the process of crystal nucleation from the liquid is important in many areas of science, from manufacture of drugs~\cite{Cox,Erdemir} and cryopreservation~\cite{Morris} to ice formation in clouds~\cite{Murray,Tang,Sosso1}. Yet, the homogeneous nucleation of crystals remains an experimentally elusive phenomenon, since it occurs at negligible rates under easily accessible, low-supercooling conditions. In particular, the detailed structure of the nascent crystal embryo completely eludes direct observation (see Refs.\,\cite{Jungblut,Sosso2} for recent reviews on the subject).

The simplest description of nucleation dates back to Volmer, Weber, and Farkas~\cite{Volmer,Farkas}, whose by now classical nucleation theory (CNT) has set the stage for the subsequent developments~\cite{Becker,Zeldovich,Frenkel,Turnbull}. Assuming the number $n$ of particles in the solid cluster as unique reaction coordinate, and using arguments from elementary thermodynamics, CNT predicts a non-monotonic cost $F_{\rm cl}(n)$ of cluster formation that correctly captures the essence of nucleation as a thermally activated phenomenon (the cluster free energy $F_{\rm cl}(n)$ attains its maximum value at the critical size $n^*$; clusters of smaller size are doomed to dissolve, while clusters larger than critical grow without bounds). In fact, the simple CNT form for $F_{\rm cl}(n)$ is far from accurate, and a number of improvements have been proposed over the years~\cite{Fisher1,Langer,Oxtoby,Dillmann,Fisher2,Prestipino1} to account, at least partially, for a diffuse solid-liquid interface and deviations of the cluster shape from spherical.

With the advent of more powerful computers in the nineties, a simulation approach to the cluster free energy has eventually become within reach~\cite{vanDuijneveldt,tenWolde1,tenWolde2,Zierenberg}. Once a criterion to identify solidlike particles is provided, Monte Carlo (MC) simulation allows, in principle, to characterize the nucleation process completely, providing access to both its rate and the structure of the nucleus (see {\it e.g.} Ref.~\cite{Valeriani} and the papers cited therein). In practice, this program is far from complete since a number of simplifications are usually adopted in the calculation, notably the idea that the size of the solid cluster is the only relevant reaction coordinate (see the follow-up discussion in Sec.\,II). This point, in conjunction with the limitations of popular path-sampling techniques in characterizing the kinetics of aggregation~\cite{Haji-Akbari3}, may explain the discrepancy in rate values often reported between simulation and experiment --- not to mention the fact that classical force fields are obviously not the same as real materials. There are other nuisances that undermine the numerical approach to nucleation, related to the partial arbitrariness in the definition of cluster particles, which make the associated free-energy profile somewhat ambiguous.

In the last few years, much attention has been paid to the process of ice nucleation from supercooled water, which has peculiarities --- owing to the directionality of hydrogen bond and to ice polymorphism --- that call for {\it ad hoc} investigation. Not for nothing, among simple substances water stands out for its wide range of metastability ($\approx 40$\,K at atmospheric pressure). Computer simulations are particularly suited to gain insight into the mechanism of ice nucleation at the microscopic scale. Indeed, a lot of work has been done already, on a variety of classical water models: ST2~\cite{Buhariwalla}, TIP4P (in its various declinations)~\cite{Espinosa2,Haji-Akbari,Zaragoza,Russo5,Quigley,Haji-Akbari4}, and monatomic water (mW)~\cite{Moore1,Espinosa2,Espinosa3,Reinhardt3,Espinosa4,Russo5,Li7,Lupi}. Also the methods employed to investigate ice nucleation have been the most varied: umbrella sampling~\cite{Buhariwalla,Reinhardt3,Russo5}, seeding approach~\cite{Espinosa2,Espinosa4,Zaragoza}, mean first passage time~\cite{Moore1}, metadynamics~\cite{Quigley}, density-functional theory~\cite{Wang}, forward-flux sampling~\cite{Haji-Akbari,Li7,Haji-Akbari4}, and aimless shooting~\cite{Lupi}. While all studies agree on the main picture ({\it i.e.}, that the nucleation cluster is polymorphic rather than structurally homogeneous), divergence persists on the critical size, nucleation rate, interface free energy, and importance of the various ice polymorphs in the first stages of crystallization.

In the present paper, I take the simpler mW model~\cite{Molinero} and focus exclusively on the ``static'' (thermodynamic and structural) aspects of ice nucleation. By considering various methods, all physically reasonable, to define solidlike particles in a simulation framework, I quantify the effect of changing the particle-classification method on the shape of the cluster free energy as a function of $n$, as well as on the structure of the nucleation cluster as it grows up to critical size. Contrary to a commonly held belief, the way of defining cluster particles has a sizable impact on many nucleation-related quantities; in particular, the relative importance of the different ice polymorphs for the structure of the nucleus varies appreciably from one protocol to the other.

The outline of the paper is as follows. In Sec.\,II I first recall the method to extract the cluster free energy from simulation, signaling advantages and shortcomings. Then, various possibilities to identify the cluster particles are described. As test-case of the numerical procedure, I examine crystal nucleation in the Lennard-Jones model, for which new benchmark data are provided. Section III is devoted to the mW model. First, I obtain novel results for its bulk properties at zero pressure, including exact free energies for all relevant ice phases (hexagonal ice, cubic ice, and ice 0~\cite{Russo5}). Next, the cluster free energy is computed, using various procedures of particle classification, and the differences between them are highlighted. As part of this investigation, I compute for each order parameter the percentage of icelike particles of various species as a function of cluster size. By critically examining the method followed, I provide an explanation for the conflicting results found. Some concluding remarks are given in Sec.\,IV.

\section{Method}
\setcounter{equation}{0}
\renewcommand{\theequation}{2.\arabic{equation}}

\subsection{General considerations}

In this section, besides presenting the method to obtain the cluster free energy in simulation, I discuss the intrinsic limitations of the procedure, then focusing on those related to the way neighbors are assigned to each particle.

A fundamental quantity in the theory of crystal nucleation is the height of the free-energy barrier (written $\Delta G^*$ in the isothermal-isobaric ensemble), defined as the difference in free energy between the liquid with and without the solid nucleus. $\Delta G^*$ enters the transition-state-theory (Kramers) expression of the rate of nucleation, proportional to $\exp\{-\beta\Delta G^*\}$, which is thought to be accurate only near coexistence, where the barrier height is much larger than $\beta^{-1}=k_BT$. Assuming a spherical cluster shape and using purely thermodynamic arguments, CNT arrives at a simple formula for the barrier height:
\be
\Delta G^*=\frac{16\pi}{3}\frac{\gamma_\infty^3}{\rho_s^2\Delta\mu^2}\,,
\label{eq2-1}
\ee
in terms of the (orientationally-averaged) free energy $\gamma_\infty$ of the planar crystal-liquid interface, the bulk-crystal number density $\rho_s$, and the supersaturation $|\Delta\mu|$. However, Eq.\,(\ref{eq2-1}) is only a rough approximation to the true barrier height, and for several reasons: 1) first of all, nucleation is a genuine non-equilibrium phenomenon, hence resort to  equilibrium thermodynamics is suspect~\cite{Chandler,Dellago}, at least far below coexistence; 2) it is by no means true that the cluster size/volume is the only relevant collective variable (CV) of nucleation, and indeed other quantities like the area of the cluster surface~\cite{Prestipino2} or the degree of crystallinity~\cite{Moroni} have been shown to be important variables as well. Furthermore, it has been argued that fluctuations in the orientational order parameter, rather than density fluctuations, play a prominent role in triggering nucleation~\cite{Russo1,Russo2}; 3) even in the assumption of a leading role for volume, the diffuseness of the cluster boundary~\cite{Burian}, deviations of the cluster from sphericity~\cite{Trudu,Prestipino1}, and the dependence of interface free energy on the crystal facet~\cite{Prestipino3} all have a non-negligible effect on the nucleation barrier and rate; 4) the other tenet, implicit in CNT, of a unique solid cluster in the mother liquid can only hold true near coexistence; 5) a final important caveat concerns the intrinsic polycrystalline nature of the nucleation cluster: even when big and post-critical, in the structure of the nucleus many different crystal orders usually coexist (see {\it e.g.} Refs.~\cite{tenWolde1,Lechner1,Lechner2,Desgranges,Mithen}).

In the simulation approach to nucleation originally put forward by ten Wolde and Frenkel~\cite{tenWolde2}, the CV chosen is the number $n$ of cluster particles, and the cluster free energy $F_{\rm cl}(n)$ is computed from the cluster-size distribution ${\cal N}_n$ ({\it i.e.}, the average number of $n$-sized clusters). The ${\cal N}_n$ distribution is generally distinct from the distribution of the largest-cluster size, see Appendix A. After a criterion to distinguish liquidlike from solidlike particles has been established, ${\cal N}_n$ can be obtained in a MC simulation using enhanced-sampling techniques~\cite{tenWolde2,Auer1,Filion,Thapar}. In the present study, I employ umbrella sampling (US) to push the size of the largest cluster forward, until slightly post-critical $n$ values are reached (see Appendix B for details on my implementation of the US method). I have performed MC runs of the mW model in the isothermal-isobaric ($NPT$) ensemble for $P=0$ and three temperatures ($T=210$\,K, 220\,K, and 230\,K), with periodic boundary conditions and cell lists, using the method of Ref.~\cite{Gelb} in order to avoid the burden of computing the maximum cluster size at each MC step. By this scheme, the maximum cluster size is computed only once every $\approx 200$ MC moves, leading to a substantial speeding up of the simulation. All simulations were carried out for samples of $N=2048$ particles, which proves to be large enough for the present purposes.

\subsection{Various methods to identify the cluster particles}

In order to pursue the objective of computing $F_{\rm cl}(n)$, a crucial step is the identification of solidlike particles for any given configuration of the sample. Obviously, there is no unique way to do this: many physically sound protocols exist, and the procedure used is ultimately a matter of choice. It is reasonable to expect that the critical size $n^*$ will (slightly) depend on this choice, but the barrier height would arguably be independent of that.

The first step in the procedure is selecting neighbors for each particle $i$. Discarding the Voronoi construction, which only comes at a high computational cost and moreover is not free of ambiguity~\cite{Saija,vanMeel}, other methods are more suited for an on-the-fly usage. In Ref.~\cite{tenWolde3}, as in most of the subsequent studies, the criterion has been adopted that two particles are neighbors to each other if their distance is less than the first minimum $r_{\rm cut}$ of the radial distribution function (RDF) of the liquid; in the following, I refer to this choice as RCUT. A different scheme is to take the neighbors of $i$ to be the $n_{\rm vic}$ particles nearest to it, as is done {\it e.g.} in Refs.\,\cite{Mickel,Mithen} (this choice will be dubbed NVIC). Another possibility is to resort to the solid-angle based nearest-neighbor (SANN) algorithm proposed by van Meel {\it et al.}~\cite{vanMeel}, which has the advantage of being parameter-free and robust to small fluctuations in the local structure (I shall refer to this prescription as SANN).

Once neighbors have been assigned to each particle, calling $\hat{\bf r}_{ij}$ the unit vector specifying the orientation of the bond between particle $i$ and its neighbor $j$, a local bond-order parameter is defined as~\cite{Steinhardt}
\be
q_{lm}(i)=\frac{1}{N_b(i)}\sum_{j=1}^{N_b(i)}Y^m_l(\hat{\bf r}_{ij})\,,
\label{eq2-2}
\ee
denoting with $N_b(i)$ the number of $i$ neighbors and with $Y^m_l$ the spherical harmonics. In a crystal, at variance with what occurs in the liquid, the $q_{lm}(i)$ add up coherently, which is the reason why these quantities are useful to identify solidlike particles (in particular, $q_{lm}$ parameters with $l=6$ are especially suited for cubic crystals). Next, following Ref.\,\cite{tenWolde2}, a normalized 13-dimensional complex vector ${\bf q}_6(i)$ is considered, with components proportional to $q_{6m}(i)$. The dot product
\be
{\bf q}_6(i)\cdot {\bf q}_6(j)=\frac{\sum_{m=-6}^6 q_{6m}(i)q_{6m}(j)^*}{\sqrt{\sum_{m=-6}^6\left|q_{6m}(i)\right|^2}\sqrt{\sum_{m=-6}^6\left|q_{6m}(j)\right|^2}}
\label{eq2-3}
\ee
measures how similar the environment of $i$ is to that of $j$. If ${\bf q}_6(i)\cdot{\bf q}_6(j)$ exceeds a certain threshold (in my case, 0.65), then particles $i$ and $j$ are said to be connected. If the number of connections is larger than a threshold $n_{\rm co}$ ({\it e.g.}, 7 for the Lennard-Jones system analyzed in Ref.\,\cite{tenWolde3}), then particle $i$ is said to be solidlike.

A major step forward eventually came with the publication of Ref.\,\cite{Lechner1}, where a useful method is proposed to distinguish between different crystal environments. While the local order parameters (OP)
\be
q_l(i)=\sqrt{\frac{4\pi}{2l+1}\sum_{m=-l}^l\left|q_{lm}(i)\right|^2}
\label{eq2-4}
\ee
are hardly able to distinguish between various crystalline phases, the modified OP
\be
\overline{q}_l(i)=\sqrt{\frac{4\pi}{2l+1}\sum_{m=-l}^l\left|\overline{q}_{lm}(i)\right|^2}\,\,\,\,\,\,{\rm with}\,\,\,\,\,\,\overline{q}_{lm}(i)=\frac{1}{N_b(i)+1}\left(q_{lm}(i)+\sum_{j=1}^{N_b(i)}q_{lm}(j)\right)
\label{eq2-5}
\ee
do that accurately, when used in combination with other parameters $\overline{w}_l(i)$, constructed from the $\overline{q}_{lm}(i)$ themselves (see Ref.\,\cite{Lechner1} for details). This happens because $\overline{q}_{lm}(i)$ keeps track of the structure of the second-neighbor shell in addition to the first shell, in a way doing the same job as the scalar product (\ref{eq2-3}). The classification of particles according to the scheme devised by Lechner and Dellago leads to a further criterion to identify cluster particles, more refined than the other three presented above, which will be denoted LD (from the initials of the two authors). Finally, it is worth mentioning the recent proposal~\cite{Piaggi} of a new method to discriminate between different crystalline orders, based on the evaluation of a local enthalpy and a local pair entropy~\cite{Giaquinta}, which promises to be a useful cheap alternative to the LD method.

Once a type (either liquidlike or solidlike) has been attached to every particle, connected assemblies of solidlike particles (clusters) are enumerated with the Hoshen-Kopelman algorithm~\cite{Hoshen}.

\subsection{Test-case: crystal nucleation in the Lennard-Jones model}

In Ref.\,\cite{Filion}, a study has been performed of the sensitivity of the cluster free energy for hard spheres to the scheme used to identify solidlike particles. Within a RCUT type of framework, the minimum number $\xi_c$ of connections required for a particle to be considered solidlike was varied from 5 to 10. Not surprisingly, it was found that the size of the nucleus depends substantially on $\xi_c$, while the barrier height is almost insensitive to it. Instead, no extensive analysis has ever been attempted (to my knowledge at least) of the dependence of the nucleation barrier on the criterion to assign neighbors to a particle. In this section, I present the results of such an analysis for the Lennard-Jones (LJ) fluid. This will serve as a reference for the subsequent study of monatomic water in Sec.\,III.

I consider the cut-and-shifted LJ model with cutoff radius 2.5~\cite{Abramo,Prestipino4} (reduced units are used throughout this section), probing the features of crystal nucleation as a function of pressure along the overcompressed-liquid branch for $T=0.90$ (see Fig.\,10 of Ref.~\cite{Abramo}). The number of particles is set equal to $N=4000$. Four values of pressure are considered, {\it i.e.}, 5.69, 6.63, 7.71, and 8.89, corresponding to number densities $\rho$ approximately equal to 0.960, 0.980, 1.000, and 1.020, respectively. For each value of $P$, as starting point of the US simulation I take the last configuration produced in a MC run of the supercooled-liquid system at the same density. I found that the liquid did not spontaneously transform into a (defective) crystalline solid within $6\times 10^5$ MC cycles (one cycle corresponding to $N$ elementary MC moves). In each US window (centered at multiples of 5 and with half-width of $\Delta n=5$), as many as $6\times 10^5$ MC cycles are produced, starting from a system configuration with a maximum-cluster size falling in the same window; averages are taken over the last $4\times 10^5$ cycles.

%
%
\begin{figure}[t]
\includegraphics[width=10cm]{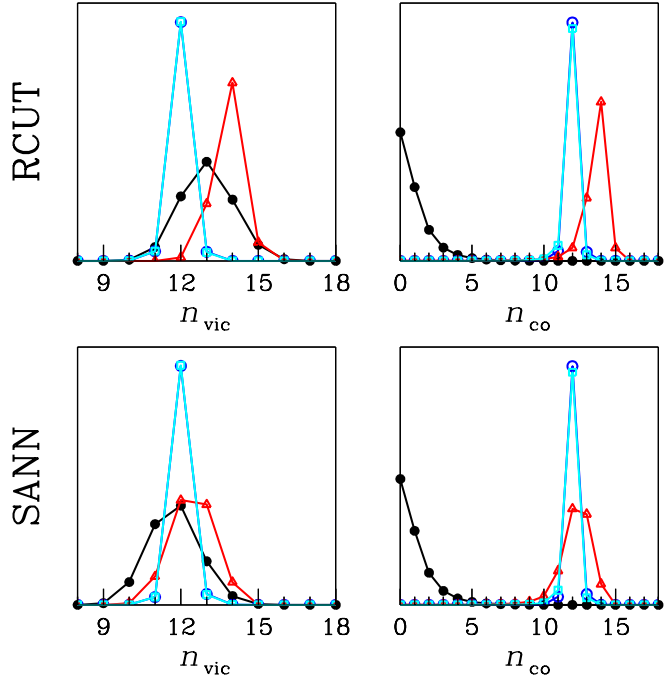}
\caption{Cut-and-shifted LJ model with $r_{\rm cut}=2.5$, for $N=4000,P=7.71$, and $T=0.90$: Distribution of the number of neighbors and of the number of connections according to the RCUT method (top) and the SANN method (bottom) in various bulk phases (supercooled liquid, black full dots; fcc, blue open dots; bcc, red triangles; and hcp, cyan squares).}
\label{fig1}
\end{figure}

I have computed $F_{\rm cl}(n)$ using the weight function (\ref{eqB-5}) and a variety of schemes for particle classification: RCUT, NVIC, SANN, and LD, see Sec.\,II.B; for $P=7.71$, other RCUT data have been generated for a different form of weight function, {\it i.e.}, Eq.\,(\ref{eqB-2}) with $\kappa=0.2$. Leaving for a moment the LD calculation aside, the following parameters have been used in the other cases: $r_{\rm cut}=1.475$ (RCUT), $n_{\rm vic}=12$ (NVIC), and $n_{\rm co}=7$ (RCUT, NVIC, SANN). In particular, $1.475$ is where the first minimum of the liquid RDF falls for $P=7.71$; $12$ is the average number of neighbors in the liquid for $P=7.71$ (when the SANN scheme is employed); finally, I have chosen $n_{\rm co}=7$, since for $P=7.71$ the number of connections is never less than 8 in any crystal, while being definitely smaller than 8 in the liquid (see Fig.\,1).

%
%
\begin{figure}
\includegraphics[width=10cm]{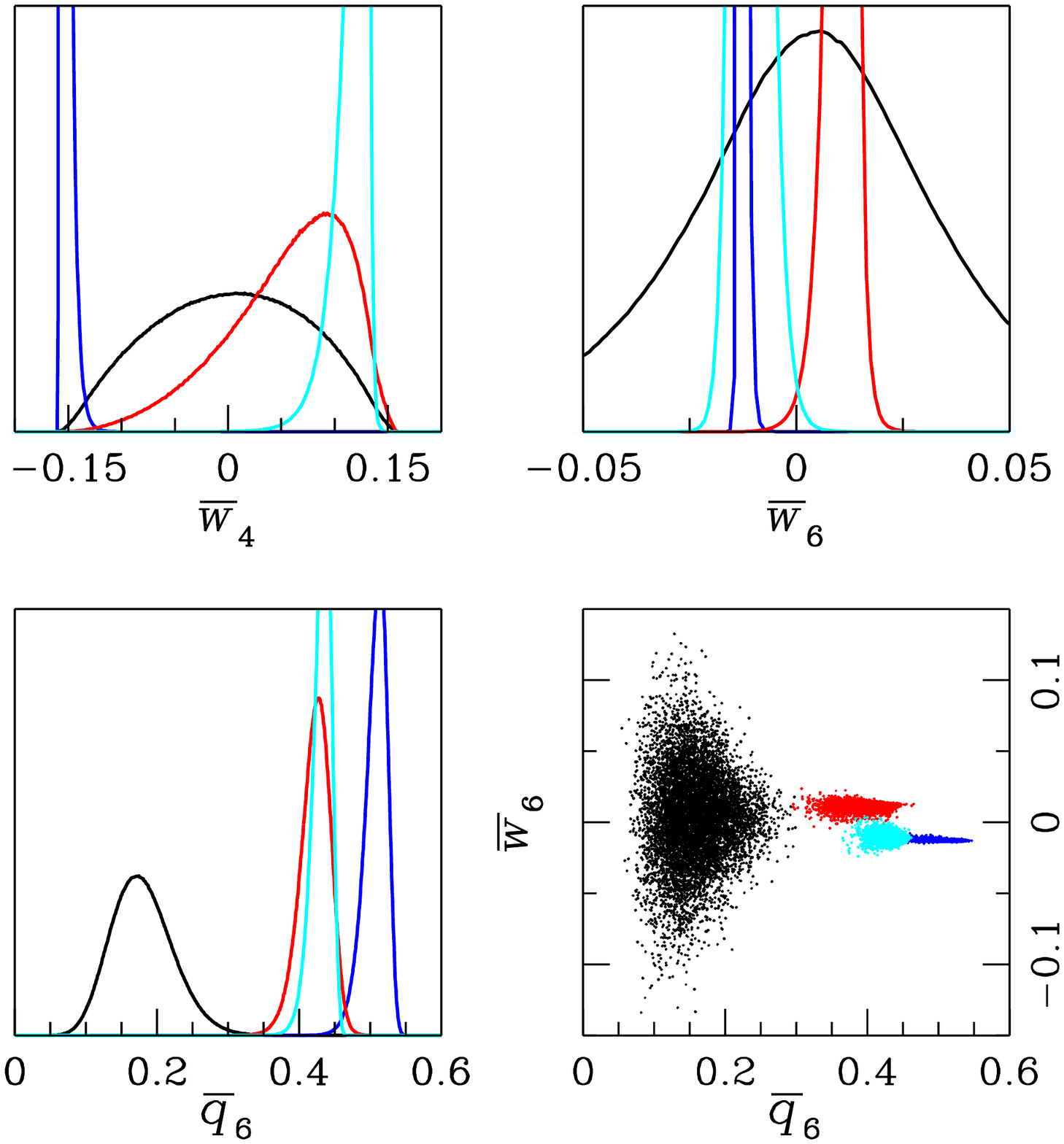}
\caption{Cut-and-shifted LJ model with $r_{\rm cut}=2.5$, for $N=4000,P=7.71$, and $T=0.90$: Orientational-OP probability distributions in various bulk phases (same color code as in Fig.\,1). In the bottom right panel, the domains of each phase are represented as scatter plots on the $\overline{q}_6$-$\overline{w}_6$ plane (each dot in the map corresponds to a particle; data are gathered from 10 evenly spaced MC configurations out of a total of $50000N$ produced at equilibrium).}
\label{fig2}
\end{figure}

%
%
\begin{figure*}
\begin{center}
\includegraphics[width=15cm]{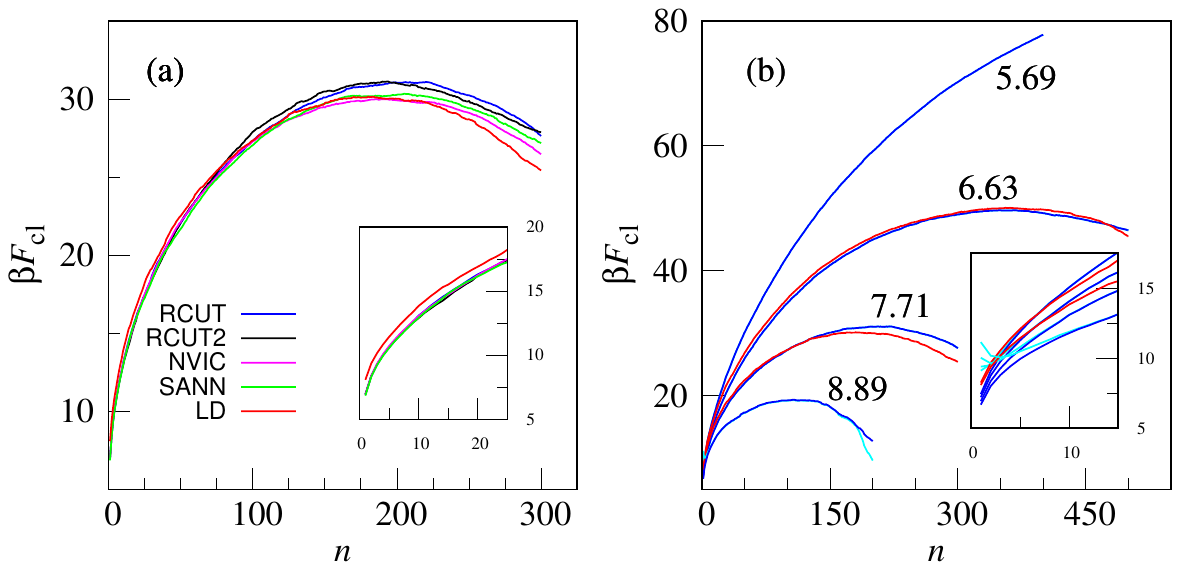}
\caption{Cut-and-shifted LJ model with $r_{\rm cut}=2.5$, for $N=4000$ and $T=0.90$: (a) The cluster free energy for $P=7.71$ as computed for a number of different schemes to identify solidlike particles (RCUT, blue; RCUT with weight function (\ref{eqB-2}) and $\kappa=0.2$, black; NVIC, magenta; SANN, green; LD, red). In the inset, a magnification of the low-$n$ region. (b) Cluster free energy for various pressures (RCUT, blue; LD, red). Cyan curves represent RCUT results for $\beta F^*_{\rm cl}(n)+\ln N$ (see Appendix A). In the inset, a magnification of the low-$n$ region.}
\label{fig3}
\end{center}
\end{figure*}

When neighbors are assigned to each particle according to, say, the RCUT scheme with $r_{\rm cut}=1.475$, but the LD method is adopted to identify the particle nature, solidlike particles can be further distinguished between fcc-like, bcc-like, and hcp-like (as usual, fcc, bcc and hcp are acronyms for face-centered-cubic, body-centered-cubic, and hexagonally-closed-packed, respectively). In the present case, particle $i$ is classified as liquidlike whenever $\overline{q}_6(i)<0.3$. For $\overline{q}_6(i)>0.3$, $i$ is bcc-like if $\overline{w}_6(i)>0$; hcp-like if $\overline{w}_6(i)<0$ and $\overline{w}_4(i)>0$; and fcc-like if $\overline{w}_6(i)<0$ and $\overline{w}_4(i)<0$. This recipe has been prepared on the basis of the OP probability distributions and order map reported in Fig.\,2, which have been obtained using RCUT for $P=7.71$ (the classification rules obtained using other schemes are practically the same).

The ensuing $F_{\rm cl}(n)$ profiles are reported in Figs.\,3(a) and 3(b). Figure 3(a) refers to $P=7.71$; besides confirming independence of $F_{\rm cl}(n)$ from the weight function employed in US simulations (see blue and black curves), we see that no important difference exists between the four neighbor schemes, at least concerning the height of the nucleation barrier (gaps between the curves are not larger than $2k_BT$ at criticality, a rather small value for a barrier $\approx 30k_BT$ high). The abscissa of the maximum, $\approx 210$, is nearly the same for all the curves, except for the LD case where the nucleus size is smaller ($\approx 180$). In the inset, a large discrepancy is seen at small $n$ between the LD free energy and the other curves, and the same also occurs for $P=6.63$ (not shown); admittedly, in the present LD scheme the density of solidlike particles for small values of $n$ is underestimated with respect to, say, the RCUT scheme.

RCUT results for all the investigated pressures are reported in Fig.\,3(b), and here contrasted with a pair of cluster free energies obtained by the LD scheme. Upon moving away from coexistence, the barrier height decreases, as it should. I have fitted the $F_{\rm cl}(n)$ profile for the three highest pressures ($6.63,7.71$, and 8.89) to the CNT form, discarding $n$ values from 1 to 20; I find that the agreement worsens with increasing pressure and that the interface free energy is nearly independent of the degree of supersaturation ($\gamma_\infty\simeq 0.56\,\epsilon/\sigma^2$). When the more accurate three-parameter Fisher-Wortis form is considered~\cite{Fisher2,Prestipino1}, the fit quality substantially improves, even though at the expenses of the precision on $\gamma_\infty$, which now lies in the range 0.6-$0.7\,\epsilon/\sigma^2$. Finally, in the inset of Fig.\,3(b) we observe the expected small-$n$ departure of $\beta F_{\rm cl}(n)$ from $\beta F^*_{\rm cl}(n)+\ln N$, cf. Appendix A, while their divergence for $P=8.89$ and $n>180$ is the clue that, eventually, crystallization becomes diffuse throughout the simulation box, and the picture of a single solid cluster ceases to be valid (as confirmed by a glance at a few system snapshots).

In Fig.\,4, we follow the evolution of the orientational OP as the nominal size $n_0$ of the maximum cluster is increased. As $n_0$ grows, each OP distribution increasingly stretches towards the values typical of the fcc crystal. It is clear from Fig.\,2 that this modification can only occur by crossing through the bcc and hcp domains, with the result that the nucleus necessarily comprises also particles of these crystalline species.

%
%
\begin{figure}
\includegraphics[width=10cm]{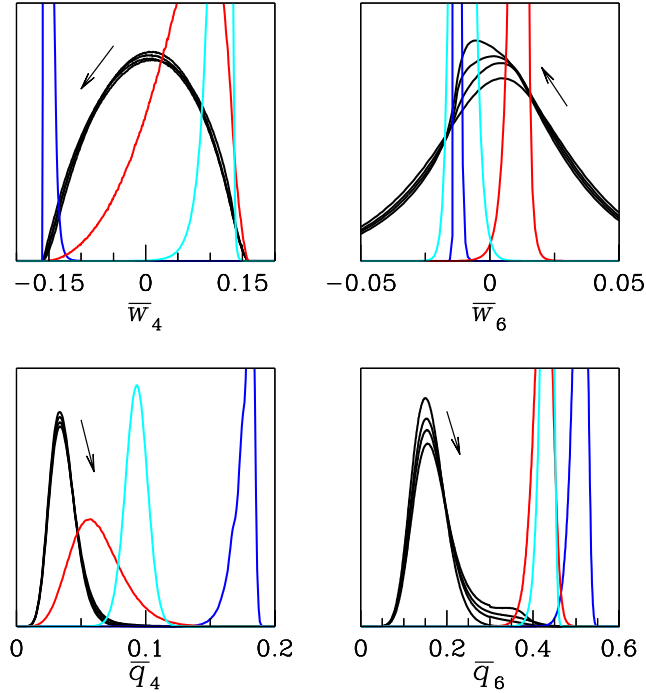}
\caption{Cut-and-shifted LJ model with $r_{\rm cut}=2.5$, for $N=4000,P=7.71$, and $T=0.90$: OP distributions for the system sampled by the US method ($n_0=5,100,200,300$, black curves --- $n_0$ grows in the direction of the arrows). Blue, red, and cyan curves are the same as in Fig.\,2. As $n_0$ grows, the distributions in the supercooled liquid change accordingly, as a result of the increasingly large number of solidlike particles present in it.}
\label{fig4}
\end{figure}

%
%
\begin{figure}
\includegraphics[width=10cm]{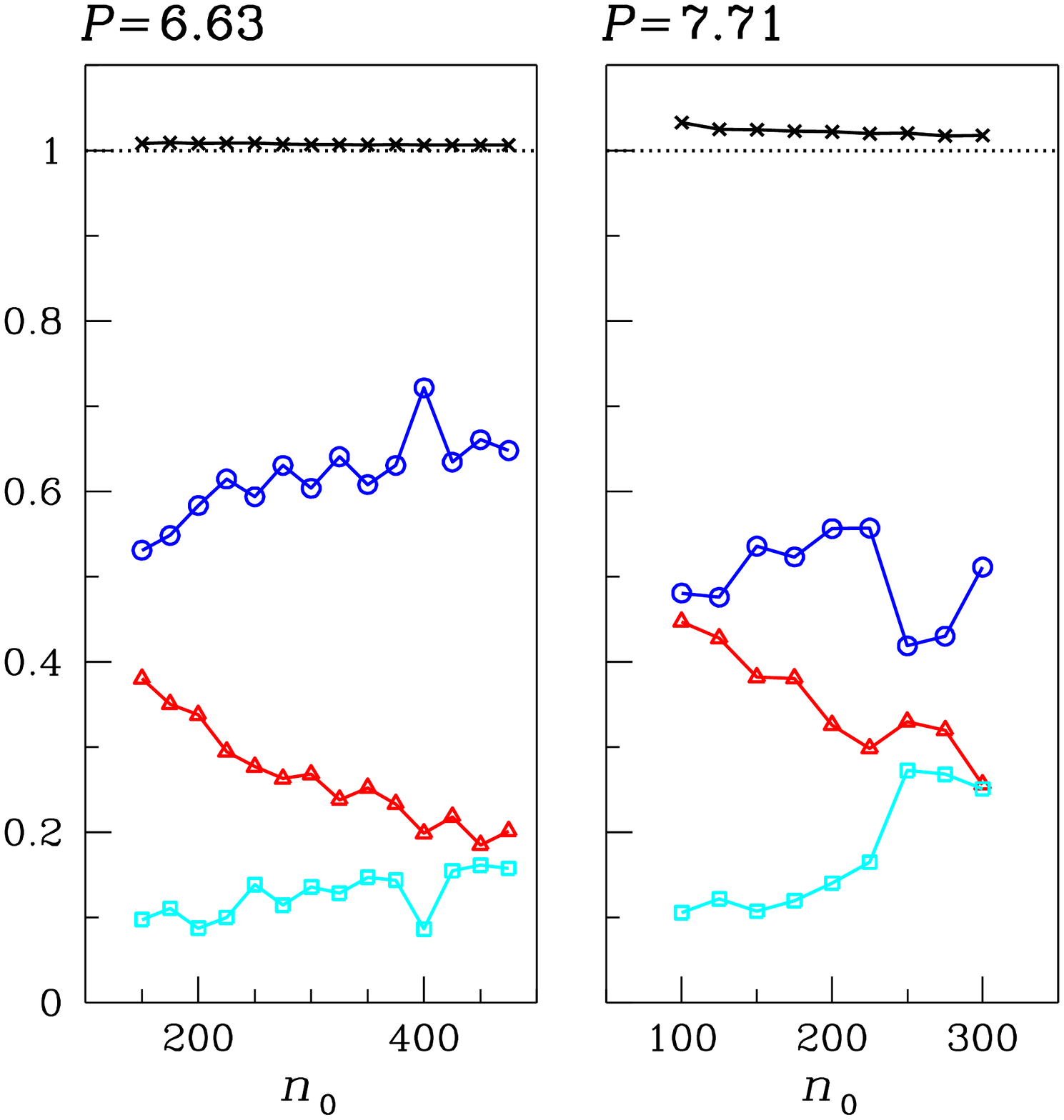}
\caption{Cut-and-shifted LJ model with $r_{\rm cut}=2.5$, for $N=4000$ and $T=0.90$: Ratio of the average number of solidlike particles in the sample to $n_0$ (same symbols and colors as in Fig.\,1), for two values of the pressure ($P=6.63$, left; $P=7.71$, right). To gather sufficient statistics, I have produced as many as $10^6$ MC cycles at equilibrium for each $n_0$. The black curve through crosses is the sum of the three colored curves; the extent to which this curve lies close to 1 indicates how much is true that all solidlike particles are gathered in a single cluster.}
\label{fig5}
\end{figure}

This is more evident in Fig.\,5, where the fractions of solidlike particles of each type are plotted for $P=6.63$ and $P=7.71$. In more precise terms, the plotted quantity is the ratio of the average number of particles, in the whole box, of each crystal species to $n_0$. We see that, for $P=6.63$, the fraction of fcc-like particles increases with $n_0$, while that of bcc-like particles decreases (hcp-like particles are only a few). Farther from coexistence ($P=7.71$), fcc-like particles are still the majority close to criticality, but in slightly post-critical clusters particles are more evenly distributed between the distinct species. Moreover, the percentage of particles not belonging to the maximum cluster is larger for $P=7.71$ than for $P=6.63$, owing to the fact that, for equal $n$, the picture of a single cluster is less valid farther from coexistence. The main message conveyed by Fig.\,5 is that precritical clusters are not tiny pieces of fcc crystal; rather, they present a high degree of polymorphism which moreover grows with pressure. Hence, polymorph selection is only accomplished in the growth regime, as also observed for softly-repulsive spheres~\cite{Desgranges2} and Gaussian particles~\cite{Mithen}. For the LJ model, the first observation of the mixed structure of clusters dates back to Ref.\,\cite{tenWolde1}, where precritical clusters were found to be predominantly bcc-like; only near the critical size, the core of the cluster became more fcc-like. This is consistent with the present findings for the cut-and-shifted LJ model, though only not too far from coexistence. Subsequent studies of LJ nucleation~\cite{Wang,Malek} have indicated that precritical clusters are rich in randomly stacked hexagonal planes, hence in hcp-like particles. Again, the same is found in the present investigation of the cut-and-shifted LJ model, where in the deep metastable region there is a large amount of hcp-like particles in the nuclei.

\section{Monatomic water}
\setcounter{equation}{0}
\renewcommand{\theequation}{2.\arabic{equation}}

\subsection{Bulk properties}

In the mW model~\cite{Molinero}, water molecules are represented as point particles interacting through suitable short-range two-body and three-body potentials, devised so as to mimic the system preference for the tetrahedral connectivity typical of the hydrogen-bond network. Starting from the Stillinger-Weber model of silicon~\cite{SW}, Molinero and Moore have changed the values of three parameters (that is, $\sigma,\epsilon$, and $\lambda$, which were set equal to $2.3925$\,\AA, $6.189$\,Kcal/mol, and $23.15$, respectively) in order to best reproduce the properties of ambient-pressure water near 273.15\,K. With the same words of Ref.\,\cite{Molinero}, structural and thermodynamic properties of water are thus reproduced ``with comparable or better accuracy than the most popular atomistic models of water, at less than 1\% of the computational cost''. For this reason, since its introduction in 2009 the mW model has been employed by many authors to assess questions relative to solidification of water at low pressure, including nucleation of ice in nanopores~\cite{Moore4,Moore6,Limmer}, in nanodrops~\cite{Amaya}, and in free-standing films~\cite{Haji-Akbari2}, waterlike anomalies~\cite{Holten3,Lu}, and the slowing down of ice nucleation induced by pressure~\cite{Espinosa3}. A recent adding is the discovery by Russo and coworkers~\cite{Russo5} of a new metastable form of ice, named ice 0, a tetragonal crystal with 12 particles in the unit cell, also observed in TIP4P water, with an alleged leading role in the crystallization of supercooled water at nearly zero pressure. In the following, I provide novel thermodynamic data also for this ice phase. To complete the picture, in a paper published few months ago~\cite{Lupi} Molinero and coworkers have challenged the idea that the preferred low-pressure solid phase of mW is hexagonal ice, suggesting that the thermodynamically stable ice in fact consists of a random sequence of cubic and hexagonal layers, with a prevalence of the former over the latter ones.

%
%
\begin{figure*}
\begin{center}
\begin{tabular}{cc}
\includegraphics[width=8cm]{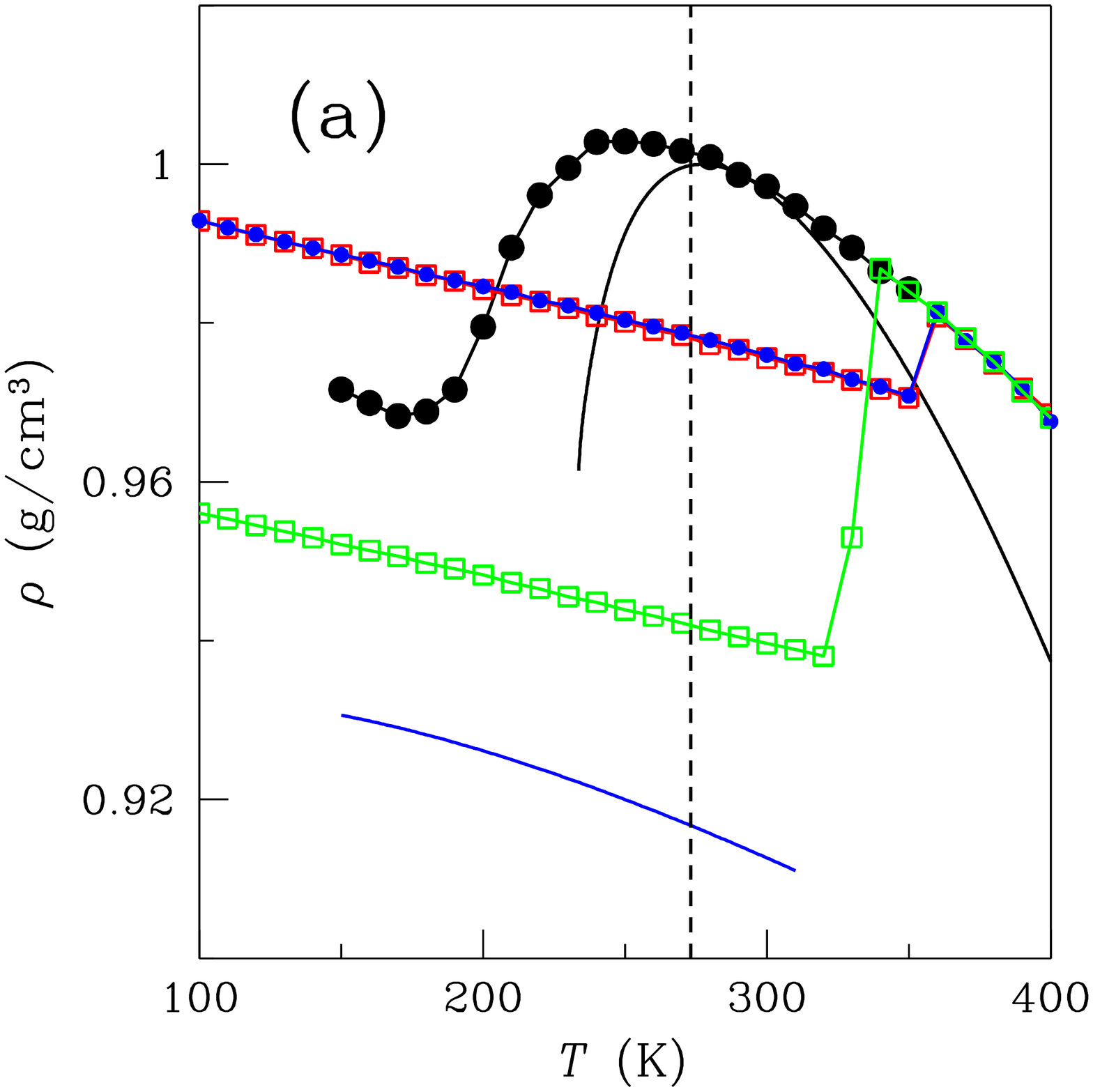} &
\includegraphics[width=8cm]{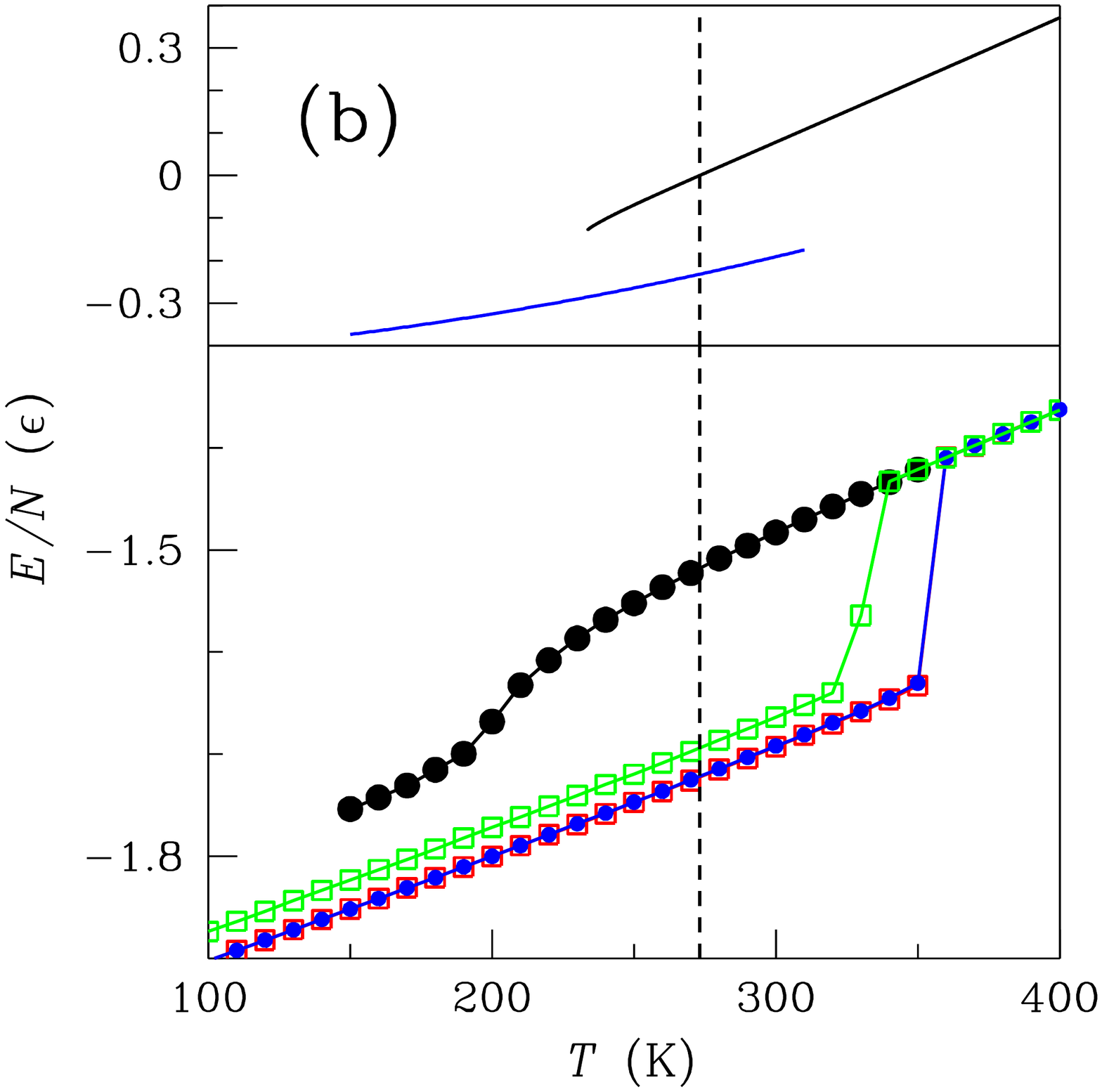}
\end{tabular}
\caption{mW model at zero pressure: (a) Mass density in g/cm$^3$, and (b) energy per particle in units of $\epsilon=6.189$\,Kcal/mol. Various phases are shown: liquid (black dots), ice I$h$ (blue dots), ice I$c$ (red squares), and ice 0 (green squares). Symbols are joined by straight-line segments to help the eye. The black and blue solid lines represent the mass density/specific energy of liquid water and hexagonal ice, respectively, as parametrized by the IAPWS-95 formulation~\cite{Wagner,Feistel}.}
\label{fig6}
\end{center}
\end{figure*}

First of all, I have tested my $NPT$ code for mW against the liquid-water density data of Moore and Molinero~\cite{Moore5}. Fixing $P=0$, I have performed simulations in a sequence, starting from the liquid at 350\,K and producing $2\times 10^5$ MC cycles for each temperature (averages are computed over the second half of the MC trajectory only). Then, I have carried out simulations of three forms of ice, starting at 150\,K and heating the sample in steps of 10\,K until 400\,K. The results, plotted in Figs.\,6(a) and (b), are fully consistent with those in Fig.\,1 of Ref.\,\cite{Moore5}. While hexagonal ice (alias ice I$h$) and cubic ice (alias ice I$c$) melted completely only at 360\,K, the ice-0 sample converted to liquid before, {\it i.e.}, at 330\,K (suggesting a weaker stability of ice 0 in comparison to ice I). Clearly, the superheating intervals can be made thinner by just lenghtening the runs or making the sample bigger (see, {\it e.g.}, Ref.~\cite{Abramo}). The properties of real water and hexagonal ice, also reported in Figs.\,6(a) and (b), are derived from the extremely accurate IAPWS-95 formulation of Wagner and coworkers~\cite{Wagner,Feistel}, see also Ref.\,\cite{Prestipino5}. The most interesting feature of Fig.\,6(a) is the density minimum at $T=170$\,K on the liquid branch. I have checked that the fraction of icelike particles (see Sec.\,III.B) is approximately 5\% at the 150\,K state point on the same branch, and that they are homogeneously distributed within the box --- hence no diffuse solidification can be called responsible for the existence of this minimum. By the way, the density minimum is not an artifact of the mW model, since a similar minimum is also found in water confined in silica nanocylinders~\cite{Liu,Mallamace}, as well as in nanoconfined fluids of softly-repulsive particles~\cite{Prestipino6,Speranza,Prestipino7}. Looking at Fig.\,6(a), the density curve of mW is not perfectly identical to that of real water, and even larger is the density gap between mW ice and real ice. Furthermore, the absolute energy scales of the mW phases are very different from the real ones (see Fig.\,6(b)). However, these are not important shortcomings of the mW model, insofar as the melting-transition properties of water and the structure of the stable ice phase at $T_m$ are well reproduced.

%
%
\begin{figure}
\includegraphics[width=15cm]{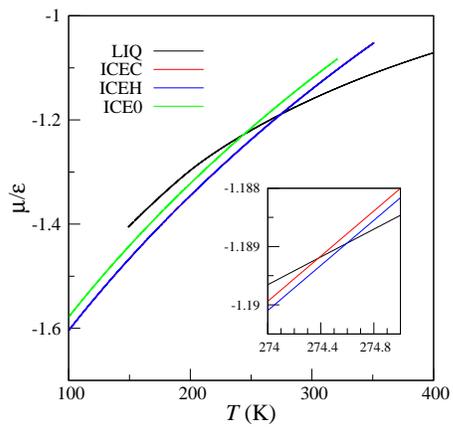}
\caption{mW model at zero pressure: Chemical potential in units of $\epsilon=6.189$\,Kcal/mol. Various phases are contrasted: liquid (black), ice I$h$ (blue), ice I$c$ (red), and ice 0 (green). Observe that the red line runs only slightly above the blue line and thus is hidden under it. In the inset, a magnification of the transition region is shown.}
\label{fig7}
\end{figure}

While the enthalpy of melting is exact by construction in mW, I have not found discussed anywhere in the literature the relative stability of the various forms of ice at zero pressure (in Ref.\,\cite{Molinero}, the higher stability of ice I$h$ over ice I$c$ has been uniquely desumed from the slightly lower melting temperature of the latter solid). Of particular relevance at low pressure are the two forms of ice I (hexagonal and cubic), and the just discovered ice-0 phase~\cite{Russo5}. I recall that the ice-I$h$ lattice for mW is hexagonal diamond (alias Lonsdaleite, or one-component wurtzite), corresponding to  the sublattice of oxygen atoms in the original ice-I$h$ crystal, whereas the ice-I$c$ lattice is just the diamond lattice. The ice-0 lattice, which is tetragonal (with $a=5.93$\,\AA\,\,and $c=10.74$\,\AA), can be reconstructed from the tabulated coordinates of Wyckoff 4b and 8i sites for the crystallographic group no.\,138 ($P4_2$/$ncm$)~\cite{BilbaoCDB}, using $x=0.84$ and $z=0.36$ (in the following, we totally neglect variations of these parameters with temperature).

I have computed the exact Helmholtz free energy for all ices at 50\,K using the Frenkel-Ladd method~\cite{FrenkelSmit,Polson}, so as to obtain the thermal evolution of the chemical potential $\mu$ for $P=0$. As for the liquid, I have assumed that the transition to ice I$h$ occurs exactly for $T_m=274.6$\,K, which is where Molinero and Moore have located it~\cite{Molinero} (in my calculation, this amounts to a $\mu$ value at melting of $-1.18894\epsilon$). We see that the stable crystalline solid is hexagonal ice at all temperatures (see Fig.\,7), even though the chemical potential of cubic ice is only just larger. Thermodynamic coexistence between ice 0 and liquid occurs near 245\,K, exactly where predicted by Russo {\em et al.}~\cite{Russo5}. I have followed the $\mu$ difference for $T=0$ between ice I$c$ and ice I$h$ as a function of pressure, and found that cubic ice becomes more stable than hexagonal ice for pressures larger than $P\simeq  1.1\times 10^5$\,atm. Moreover, ice 0 is always less stable than ice I, but more stable than ice II~\cite{ice2} or III~\cite{ice3}. Clearly, the accuracy of mW as a realistic model of water rapidly worsens upon increasingly departing from standard conditions.

\subsection{Barrier to ice formation}

The current lower limit for metastability of water at ambient pressure is 232\,K~\cite{Mossop}, which gives an upper estimate of the so-called homogeneous nucleation temperature $T_{\rm H}$ (see Appendix B). Early computer simulations of mW at zero pressure~\cite{Moore1} have shown that the crystallization rate reaches a maximum around 202\,K (to be contrasted with $225$\,K for water), below which solidification occurred so rapidly that the liquid did not have the time to equilibrate. This value of $T_{\rm H}$ was later confirmed by other authors~\cite{Li,Li7,Haji-Akbari2}. Reinhardt and Doye have reconstructed by umbrella-sampling simulations the nucleation barrier of mW at $P=1$ bar and $T=220$\,K, finding a height of $23\,k_BT$ and a critical size of 110~\cite{Reinhardt3}. More recently, Russo {\em et al.}~\cite{Russo5} have computed the nucleation free energy for $P=0$ and $T=215.2$\,K, finding a height of approximately $23\,k_BT$ and a nucleus of 60 particles. However, other authors have found values of the critical size that are much larger. For instance, Moore and Molinero report in \cite{Moore1} that ice nuclei near 210\,K consist of about 120 particles, while Li and coworkers~\cite{Li7}, using forward-flux sampling, find at 220\,K a crystal nucleus of 265 particles. Finally, Lupi {\it et al.} report critical sizes in the range 240-580 for a few cluster-size variables all well performing as one-dimensional reaction coordinates~\cite{Lupi}. This wide range of values for the nucleus size can be explained by the different status attributed by different authors to interfacial particles, and is in part due also to the different ability of umbrella sampling and forward-flux sampling to relax the local structure of a dense system. As to the structure of ice nuclei, many authors agree on the fact that they contain both cubic and hexagonal ice in roughly equal proportions~\cite{Reinhardt3,Haji-Akbari,Haji-Akbari4,Li7,Lupi}, while Russo {\it et al.} observe pre-critical clusters to consist mostly of ice-0 particles, which are converted on growth --- from the core outwards --- into cubic ice.

I have carried out US simulations of mW at three temperatures, $T=210$\,K, 220\,K, and 230\,K, for $P=0$ and $N=2048$. In the aim to determine to what degree the cluster free energy and the composition of the nucleus are sensitive to the particle-classification method, I have employed three different LD schemes (see Sec.\,II.B), which are built on the basis of the $\overline{q}_4$ and $\overline{q}_6$ distributions in the bulk phases, for $T=220$\,K (no significant change occurs if $230$\,K is chosen instead). In the first scheme (LD-A), used in tandem with RCUT, we argue as if the ice-0 phase did not exist. The cutoff distance is set equal to $r_{\rm cut}=1.43\,\sigma=3.42$\,\AA, so as to encompass the first coordination shell of any ice (Sanz and coworkers~\cite{Sanz4} have used a similar prescription in their study of ice nucleation in the TIP4P model by the seeding approach). Looking at Fig.\,8(a), particle $i$ is classified as liquidlike when $\overline{q}_6(i)<0.415$ and icelike when $\overline{q}_6(i)>0.415$; an icelike particle is of hexagonal type if $\overline{q}_4(i)<0.425$, while being of cubic type if $\overline{q}_4(i)>0.425$. We see in Fig.\,8(a) that the (bimodal) scatter plot of ice 0 lies completely over the liquid one, meaning that the LD-A scheme is unable to distinguish an ice-0 particle from a liquidlike particle.

%
%
\begin{figure*}
\begin{center}
\begin{tabular}{ccc}
\includegraphics[width=5cm]{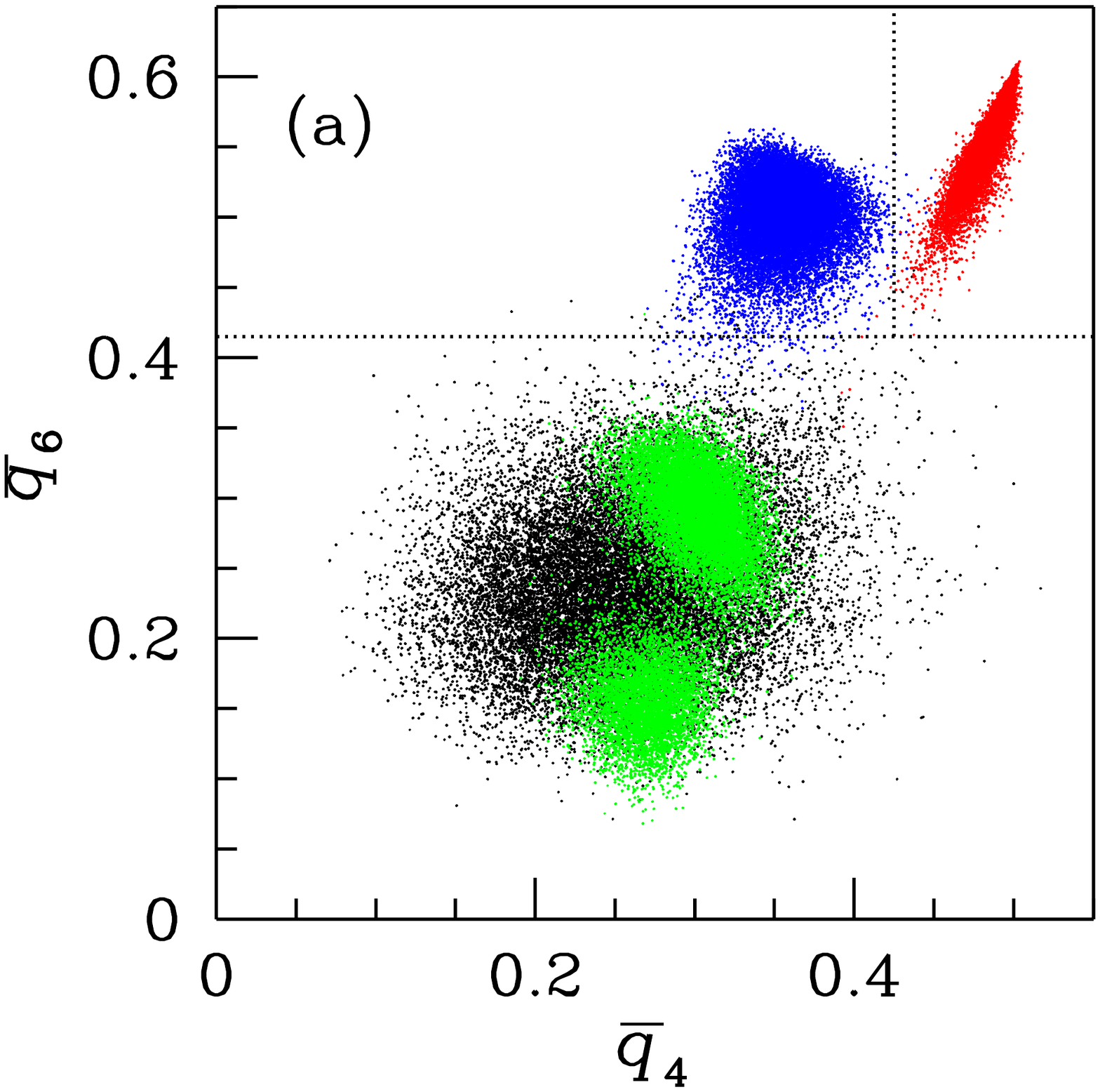} &
\includegraphics[width=5cm]{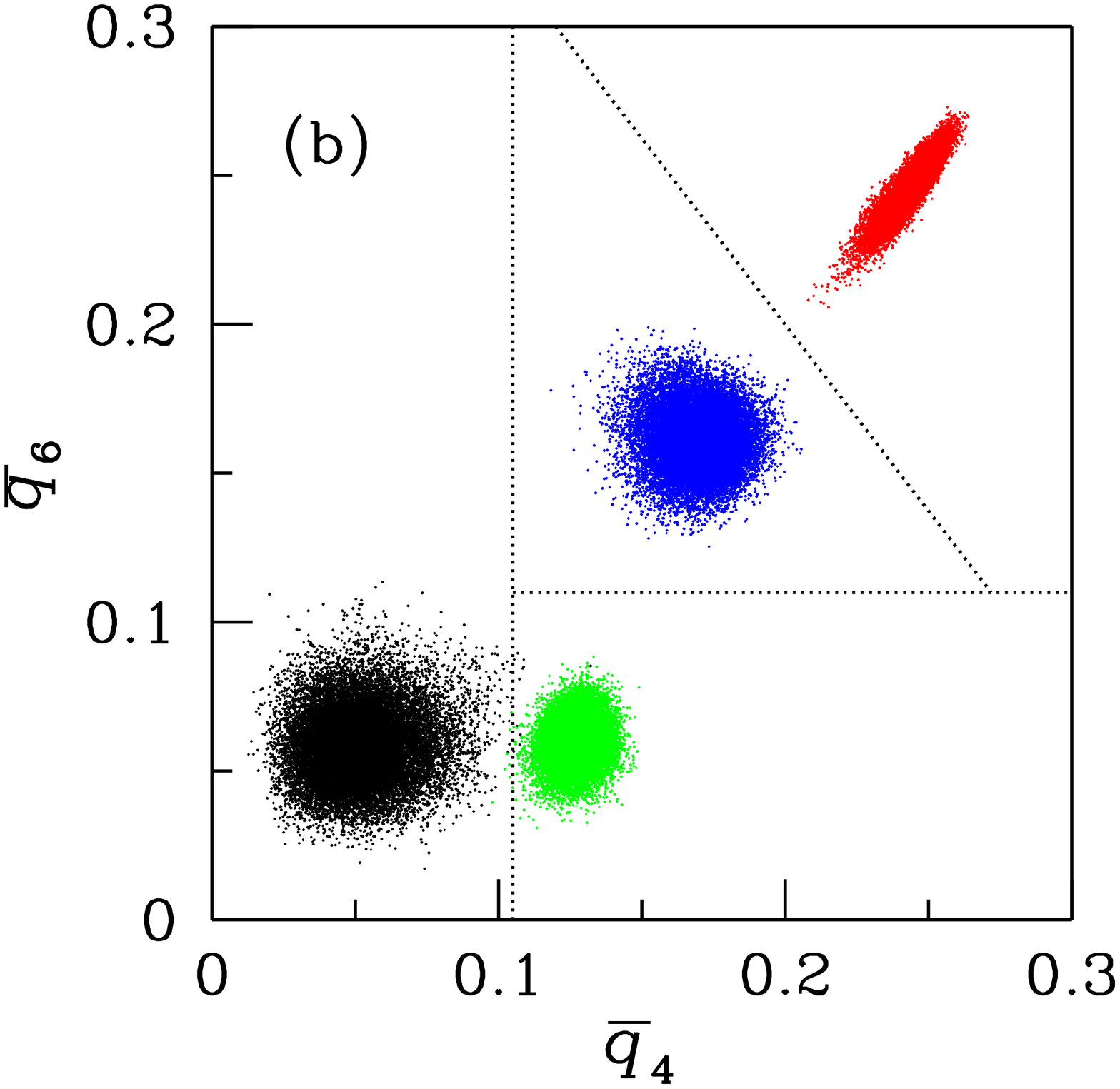} &
\includegraphics[width=5cm]{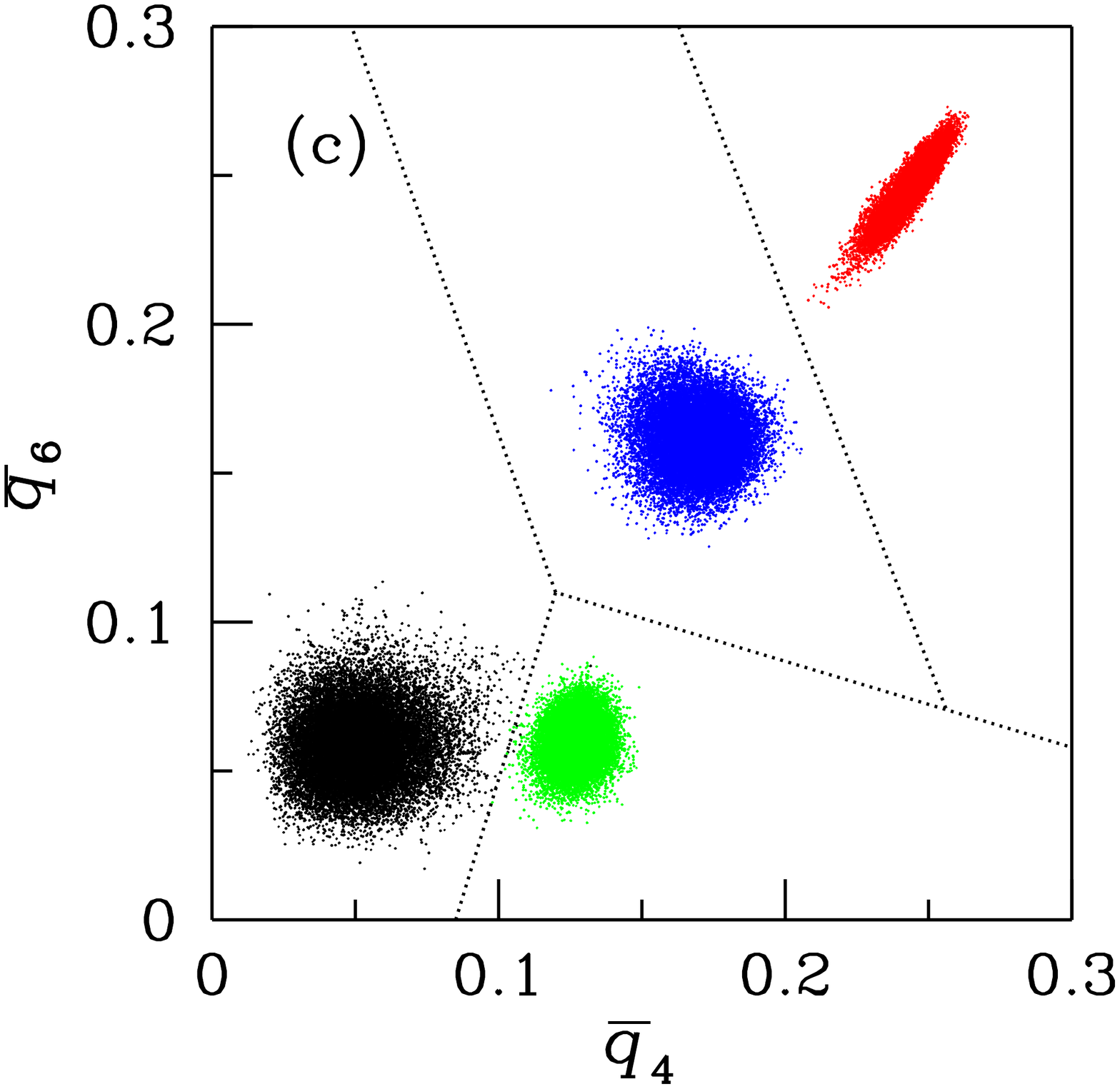}
\end{tabular}
\caption{mW model for $P=0$ and $T=220$\,K: Scatter plots on the $\overline{q}_4$-$\overline{q}_6$ plane for various bulk phases (supercooled liquid, black; ice I$h$, blue; ice I$c$, red; and ice 0, green). Each dot in the map corresponds to a particle; data are gathered from 10 evenly spaced MC configurations out of a total of $100000N$ produced at equilibrium). Each panel is relative to a different particle-classification scheme: (a) LD-A; (b) LD-B; (c) LD-C. The dotted lines in each panel correspond to the assumed subdivision of the OP plane in phase regions.}
\label{fig8}
\end{center}
\end{figure*}

For the other two schemes, the LD method is used on top of NVIC, for $n_{\rm vic}=16$ (indeed, there are 16 particles, on average, in the first two coordination shells of any ice). In the LD-B scheme, see Fig.\,8(b), particle $i$ is dubbed liquidlike when $\overline{q}_4(i)<0.105$ and icelike when $\overline{q}_4(i)>0.105$. In turn, an icelike particle is of ice-0 type if $\overline{q}_6(i)<0.11$; otherwise, it is of either hexagonal or cubic type whether it is true or not that $\overline{q}_4(i)/0.36+\overline{q}_6(i)/0.45<1$. In developing the LD-C scheme, I take greater care in dividing the $\overline{q}_4$-$\overline{q}_6$ plane into phase regions, see Fig.\,8(c), and consider particle $i$ to be liquidlike when $0.11+22/7(\overline{q}_4(i)-0.12)<\overline{q}_6(i)<0.11-8/3(\overline{q}_4(i)-0.12)$. Otherwise, $i$ is an icelike particle: of ice-0 type, if $\overline{q}_6(i)<0.11-11/38(\overline{q}_4(i)-0.12)$; instead, when $\overline{q}_6(i)>0.11-11/38(\overline{q}_4(i)-0.12)$, $i$ is of either hexagonal or cubic type whether $\overline{q}_4(i)/0.285+\overline{q}_6(i)/0.7<1$ or not. Notice that the above classification rules are different from those considered in Refs.\,\cite{Russo5,Espinosa3,Reinhardt3}: Russo and coworkers have implemented a criterion of classification based on the calculation of $\overline{q}_4$ and $\overline{w}_4$, whereas Espinosa {\it et al.} have chosen the same parameters $\overline{q}_4$ and $\overline{q}_6$ as employed here, but resorting to the RCUT scheme for the selection of neighbors. Yet different is the method adopted by Reinhardt and Doye, which have used $q_3$ to define a connectivity measure capable to distinguish icelike from liquidlike particles.

%
%
\begin{figure}
\includegraphics[width=10cm]{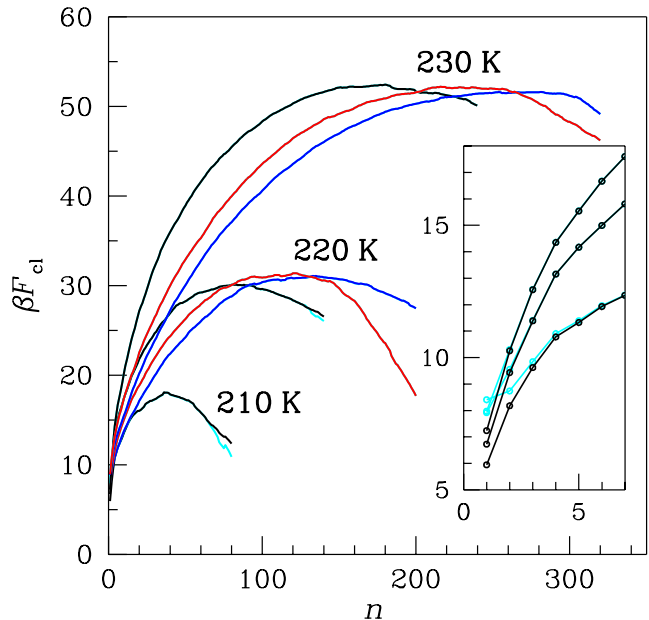}
\caption{mW model at zero pressure: Cluster free energy in units of $k_BT$, $\beta F_{\rm cl}(n)$, at three temperatures, $T=210$\,K, 220\,K, and 230\,K, as computed through different schemes for the classification of icelike particles (LD-A, black; LD-B, blue; LD-C, red). Cyan curves represent $\beta F^*_{\rm cl}(n)+\ln N$ (same color for all schemes). In the inset, a magnification of the low-$n$ region for the LD-A case only.}
\label{fig9}
\end{figure}

The profile of $F_{\rm cl}(n)$ is reported in Fig.\,9 for three values of $T$ (for $T=210$\,K, the calculation was carried out for LD-A only). As is evident from the picture, different particle-classification schemes lead to cluster free energies that are rather different, more than in the LJ case examined in Sec.\,II.C. Especially, the nucleus size varies much from one scheme to the other (for instance, for $T=220$\,K the critical size is about 80 for LD-A, 135 for LD-B, and 120 for LD-C; even larger gaps are found for $T=230$\,K). On the contrary, the barrier-height estimates are the same to within 2$k_BT$ ($\approx 30k_BT$, for $T=220$\,K), as similarly found in Ref.\,\cite{Lupi} for the free-energy barriers relative to various cluster-size coordinates. Notice that the definition of nucleation barrier adopted here is different from that considered in Refs.\,\cite{Russo5,Reinhardt3}, where it is rather $F_{\rm cl}(n^*)-F_{\rm cl}(1)$ that is called barrier height (were the same convention be used, the barrier height for $T=220$\,K will be close to that reported by Reinhardt and Doye). Moreover, no diffuse crystallization occurred spontaneously in the supercooled liquid for $T=210$\,K, at least within $1.5\times 10^6$ MC cycles and provided that LD-A is employed (I have also tried the LD-B scheme, but the sample melted during the run, implying that $T_{\rm H}$ shows a dependence on the simulation protocol). Starting from slightly above the critical size, the LD-A free energy of the maximum cluster $F_{\rm cl}^*$ departs, for both 210\,K and 220\,K, from the cluster free energy $F_{\rm cl}$, signaling a diffuse crystallization throughout the box ({\it i.e.}, the simulated sample has entered the growth regime).

Then, I have estimated the interface free energy of mW at 220\,K, by fitting the cluster free energy to the simple CNT law, finding for all schemes $\gamma_\infty\simeq 42$\,mJ/m$^2$, a value substantially larger than the one predicted by Reinhardt and Doye (26\,mJ/m$^2$)~\cite{Reinhardt3} and those found by others (28-31\,mJ/m$^2$)~\cite{Li7,Limmer,Gianetti}. I have verified that this discrepancy is entirely due to the fitting procedure, at least when the LD-A scheme is assumed. Indeed, when a nonzero Tolman's delta and an offset are included, the optimal value of $\gamma_\infty$ within LD-A significantly decreases, reaching 30\,mJ/m$^2$ at 220\,K, while remaining definitely larger for LD-B and LD-C. The bad performance of the latter two schemes in predicting the correct value of $\gamma_\infty$ casts a shadow on their effectiveness as adequate schemes for nucleation (see more below).

A further comment is in order. In CNT, the height $\Delta G^*$ of the nucleation barrier and the critical size $n^*$ are related by $\Delta G^*=(1/2)|\Delta\mu|n^*$. On account of this formula, LD-C would be perfect for both $T=220$\,K (where $\beta|\Delta\mu|=0.536$) and $T=230$\,K (where $\beta|\Delta\mu|=0.428$), while LD-A (LD-B) leads to a $n^*$ value that is systematically smaller (larger) than expected from the same equation. However, I argue that this result is of little value, since (a) CNT is actually inaccurate as a theory for $\Delta G(n)$, as widely recognized; (b) the value of $n^*$ depends strongly on the details of particle classification, much more than the barrier height. Therefore, LD-C could not be considered superior to LD-A just on the basis of the performance in the above test. The same concern about the accuracy of CNT may also be expressed for any method of computing the nucleation rate which deeply relies on this theory, such as the seeding method.

%
%
\begin{figure*}
\begin{center}
\begin{tabular}{ccc}
\includegraphics[width=5cm]{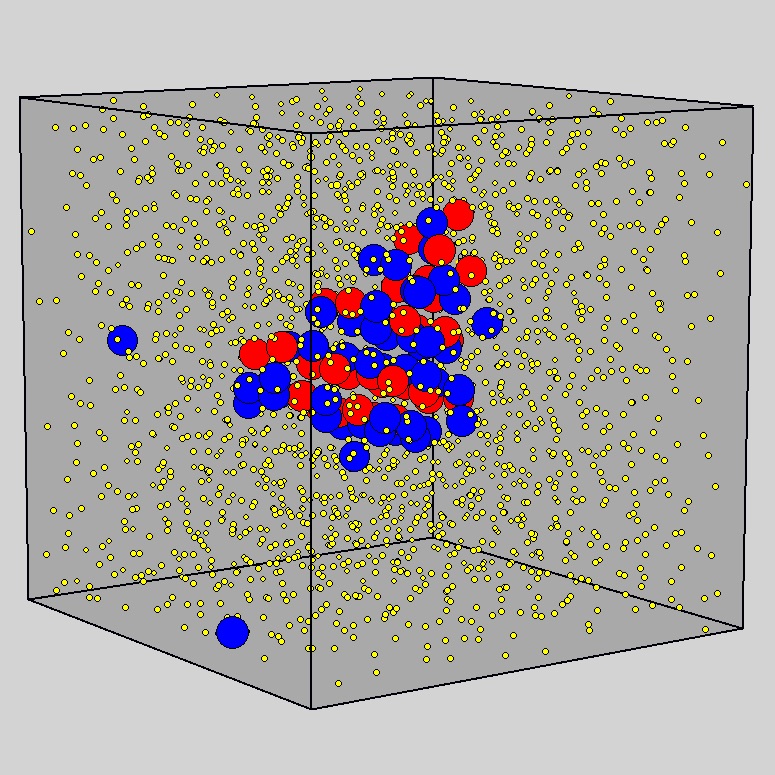} &
\includegraphics[width=5cm]{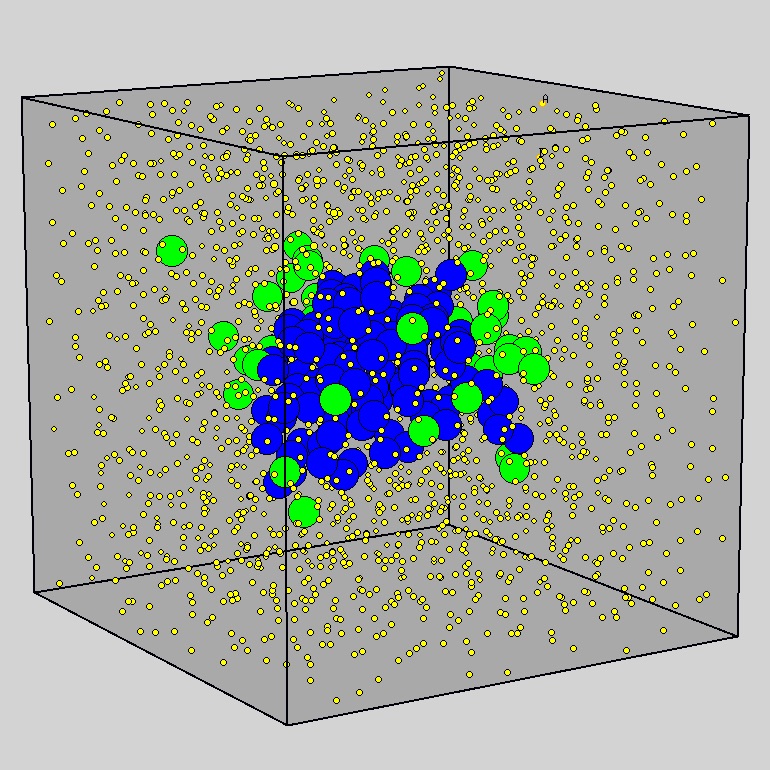} &
\includegraphics[width=5cm]{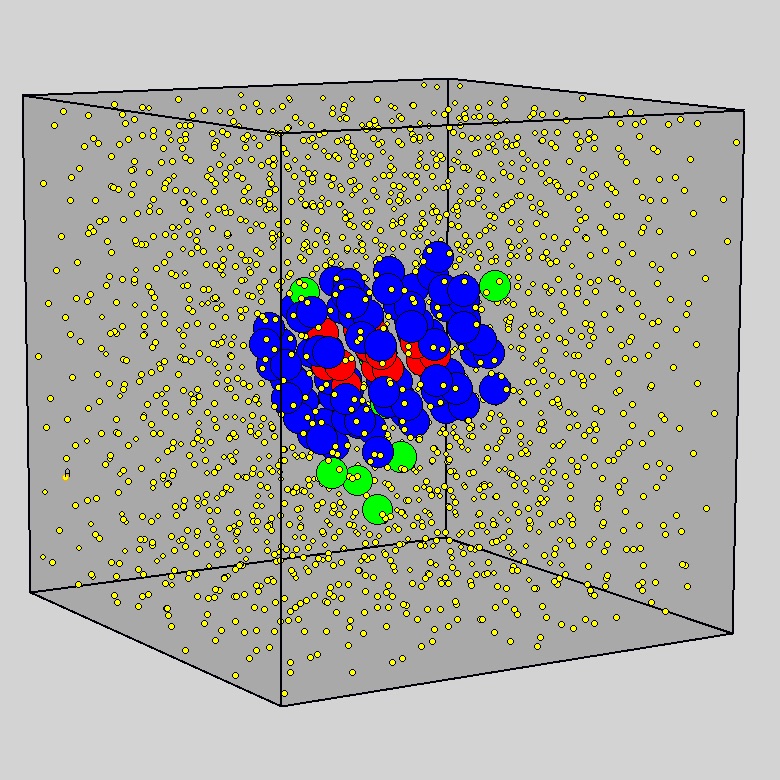}
\end{tabular}
\caption{mW model for $P=0$ and $T=220$\,K: Typical configuration near the barrier top of the supercooled-liquid sample with a solid cluster inside (for the sake of clarity, the center of mass of the maximum cluster has been moved to the center of the simulation box). The configurations are taken from US simulations carried out by employing a specific particle-classification scheme: LD-A ($n=84$, left); LD-B ($n=139$, center); LD-C ($n=118$, right). Icelike particles are represented as spheres of diameter $\sigma$. Different colors refer to different types of particle: liquid (yellow dots of diameter $0.2\sigma$); ice I$h$ (blue dots); ice I$c$ (red dots); ice 0 (green dots).}
\label{fig10}
\end{center}
\end{figure*}

%
%
\begin{figure*}
\begin{center}
\begin{tabular}{ccc}
\includegraphics[width=5cm]{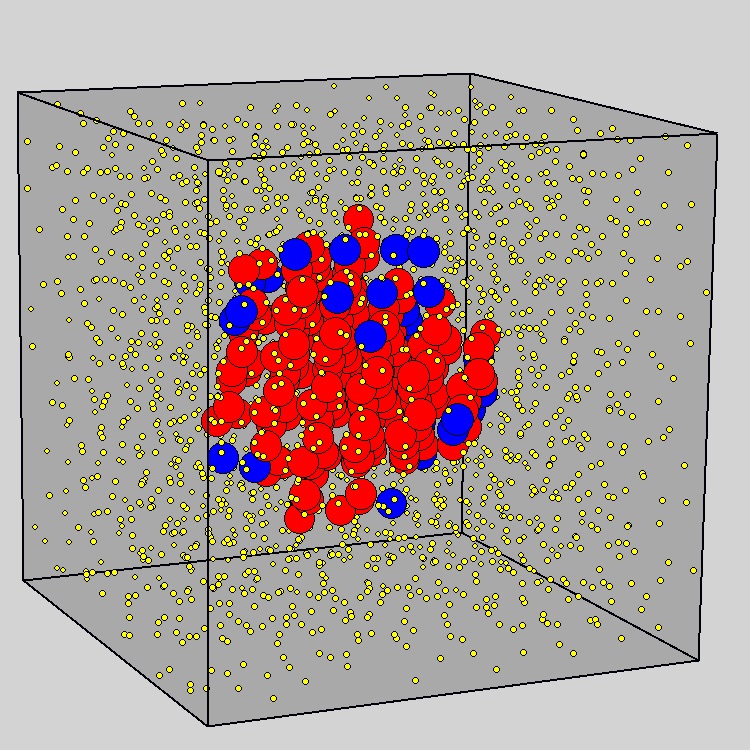} &
\includegraphics[width=5cm]{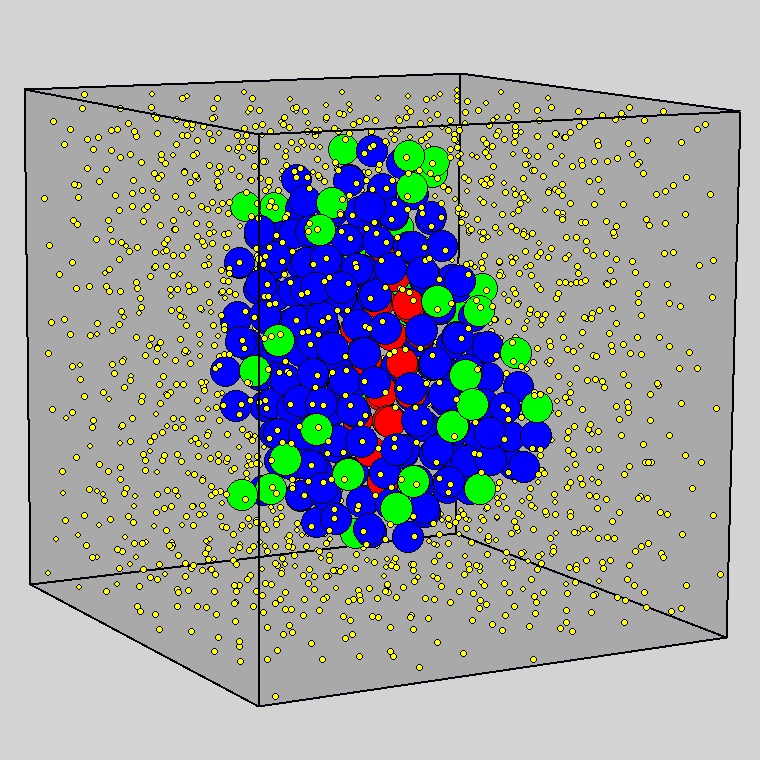} &
\includegraphics[width=5cm]{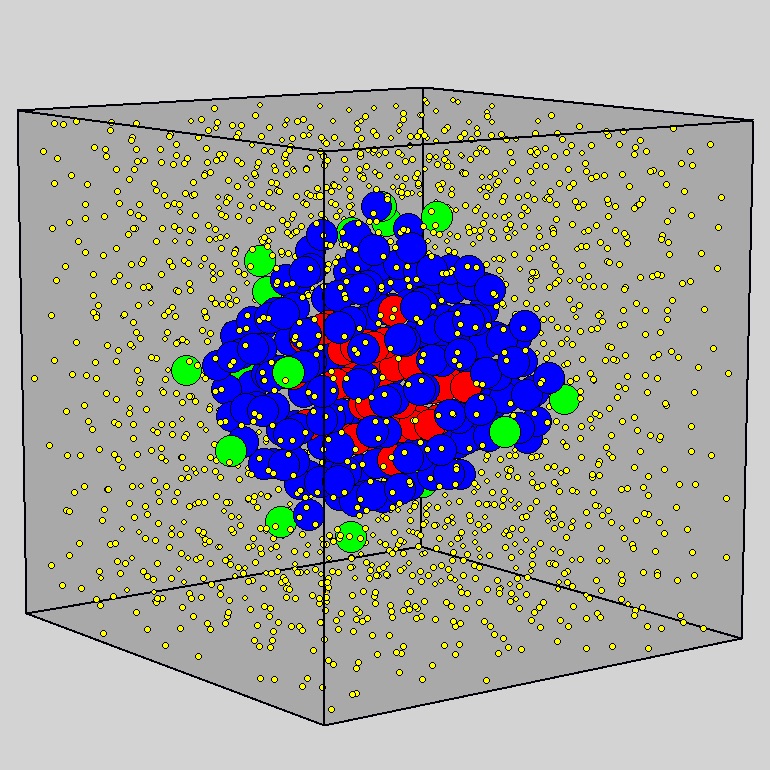}
\end{tabular}
\caption{mW model for $P=0$ and $T=230$\,K: Typical configuration near the barrier top of the supercooled-liquid sample with a solid cluster inside. The configurations are taken from US simulations carried out by employing a specific particle-classification scheme: LD-A ($n=179$, left); LD-B ($n=283$, center); LD-C ($n=237$, right). Same notation and symbols as in Fig.\,10.}
\label{fig11}
\end{center}
\end{figure*}

%
%
\begin{figure*}
\begin{center}
\begin{tabular}{ccc}
\includegraphics[width=5cm]{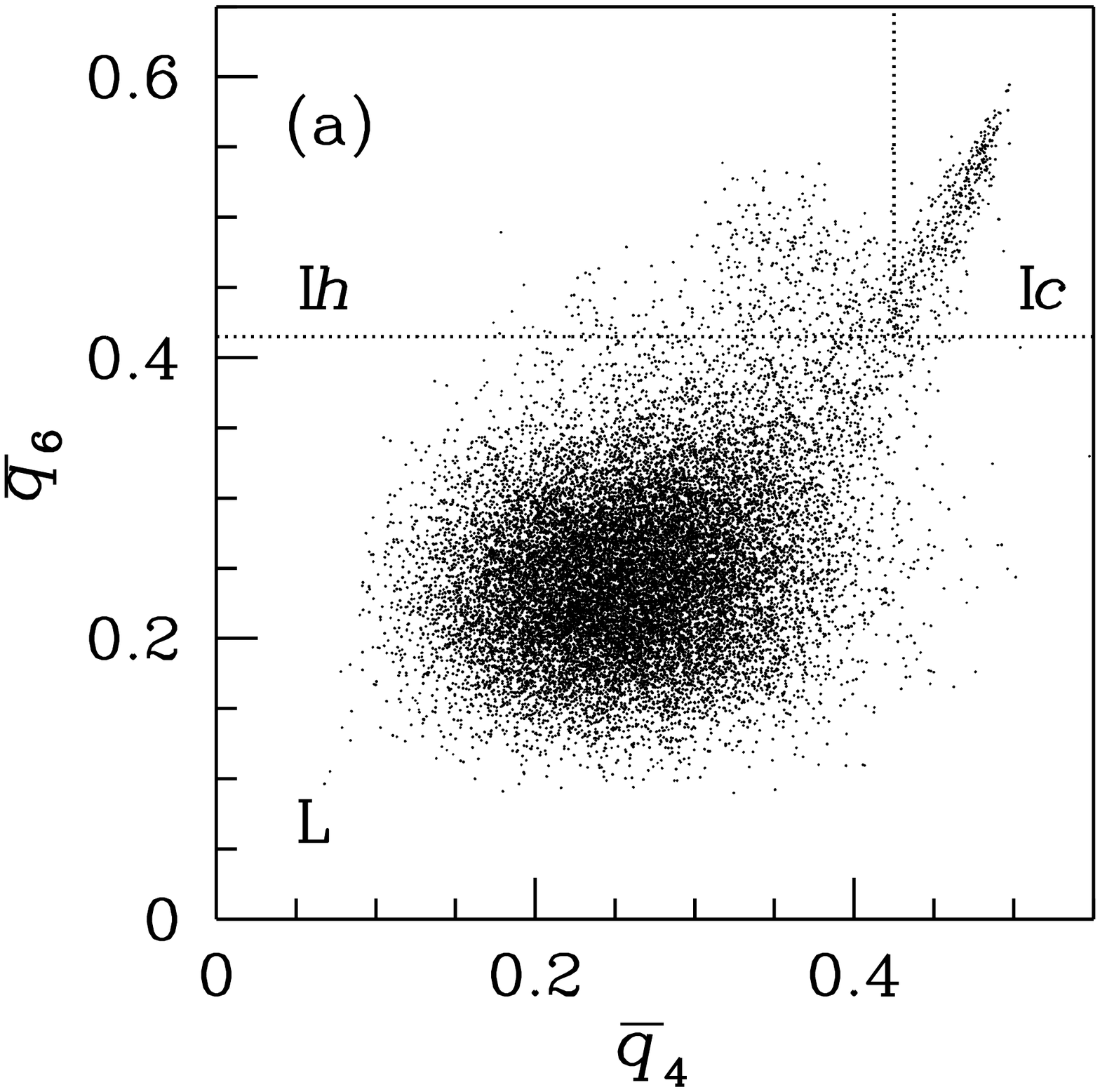} &
\includegraphics[width=5cm]{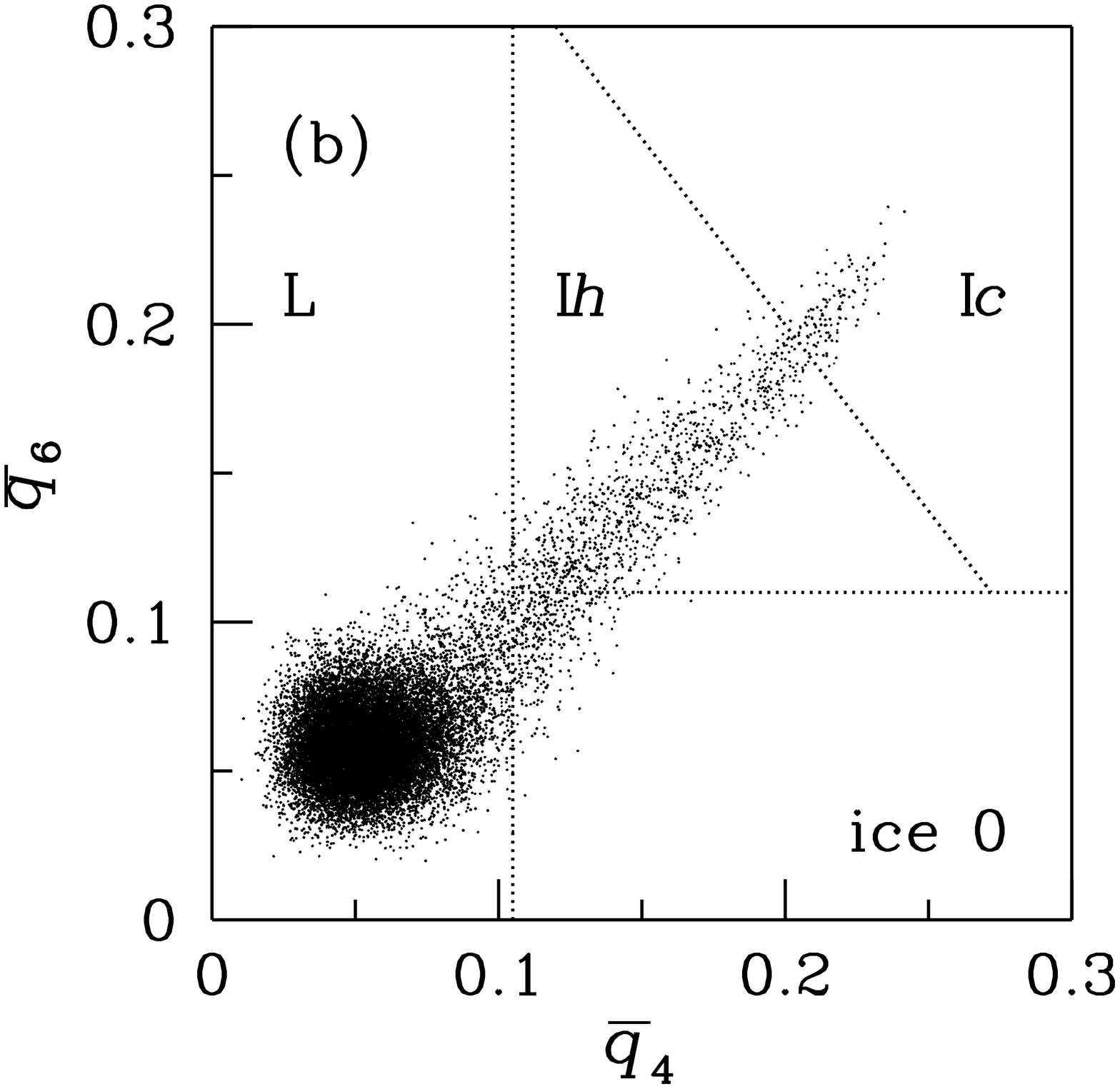} &
\includegraphics[width=5cm]{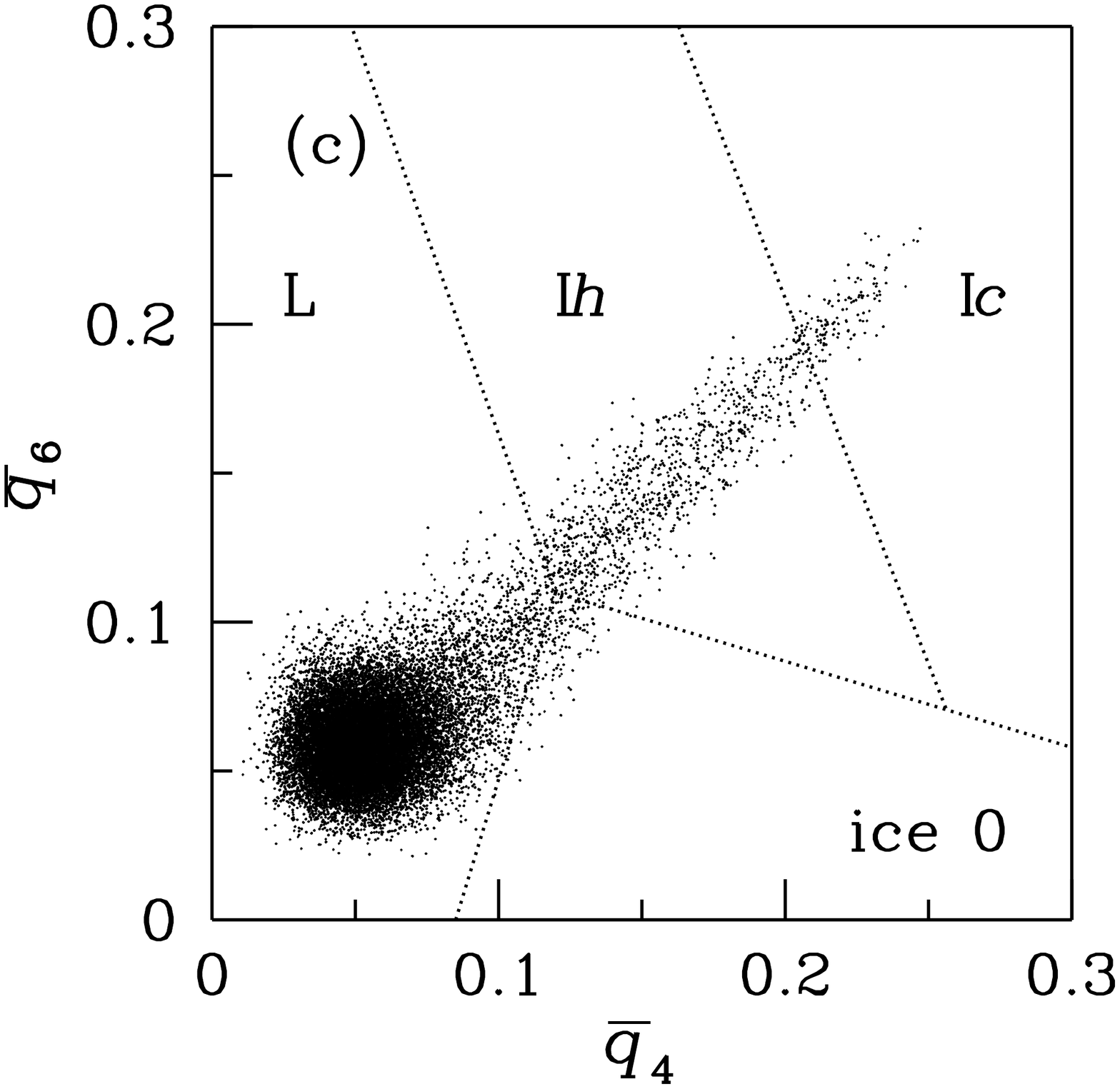}
\end{tabular}
\caption{mW model for $P=0$ and $T=220$\,K: Scatter plots on the $\overline{q}_4$-$\overline{q}_6$ plane for the sample simulated by the US method near the barrier top. Each panel is relative to a different particle-classification scheme: (a) LD-A, $n_0=80$; (b) LD-B, $n_0=136$; (c) LD-C, $n_0=120$. A single plotted dot corresponds to a particle; data are gathered from 10 evenly spaced MC configurations out of a total of $800000N$ produced at equilibrium. The phase boundaries (dotted lines) are the same as plotted in Figs.\,8(a)-(c).}
\label{fig12}
\end{center}
\end{figure*}

Looking at typical particle configurations near the top of the nucleation barrier (Figs.\,10 and 11), it stands out immediately the difference in nucleus structure between LD-A and the other schemes. In LD-A, the amount of cubic-ice particles is comparable or even larger than the number of hexagonal-ice particles, notwithstanding ice I$c$ is (marginally) less stable than ice I$h$. For LD-B and LD-C, hexagonal ice is instead dominant in the nucleus structure, with only a few cubic-ice and ice-0 particles relegated, respectively, to the core and to the surface of the nucleus (the same also occurs for clusters substantially smaller/larger than critical). In this regard, it is worth observing that the nuclei obtained at 230 K are much smaller than that (435) computed in Ref.\,\cite{Li7}, or that reported in Ref.\,\cite{Lupi} (450), which may cast doubts on the adequateness of a sample of 2048 particles for the present investigation. However, I hardly see how this large difference in critical size can be imputed to finite-size effects, considering that a discrepancy of a similar magnitude occurs at 220 K, where Ref.\,\cite{Li7} reports $n^*=265$, that is, much larger than the $n^*$ values of Fig.\,10, as well as that ($\approx 110$) found in Ref.\,\cite{Reinhardt3}. Again, a possible way out of this conundrum is to observe that the critical size is strongly sensitive to how solidlike particles are precisely identified and classified.

The hints coming from the system snapshots find full confirmation in the shape of the scatter plot at the barrier top for, say, $T=220$\,K --- see Fig.\,12. Here we observe how the scatter plot on the $\overline{q}_4$-$\overline{q}_6$ plane is modified, compared to the pure supercooled liquid, when the system is brought by the US method up to the nucleus size $n^*$. In Fig.\,12(a), the case of LD-A is illustrated. We see a clear preference for nuclei rich in cubic-ice particles, which may just reveal that the chosen $r_{\rm cut}$ is too small to allow ice I$h$ prevail locally on ice I$c$. However, a more convincing explanation is the following: Cubic ice is not so heavily unfavored with respect to hexagonal ice, hence LD-A just highlights the possibility that, in a disordered (liquid) environment, the formation of ice occurs along a path through a cubic-rich, or even fully cubic nucleus (see Fig.\,11a). As mentioned before, the relevance of these nuclei to water crystallization has been underlined many times in the literature~\cite{Haji-Akbari,Zaragoza,Malkin,Amaya,Li7,Lupi,Moore2,Reinhardt3}. In particular, it has been argued in Ref.\,\cite{Lupi} that the preference of mW for nuclei of mixed character arises from the large free-energy gain associated with random mixing of cubic and hexagonal layers.

On the other hand, the scatter plots for LD-B and LD-C look as expected on the basis of the higher stability in bulk of hexagonal ice over cubic ice, see Figs.\,12(b) and (c). As the crystallite gets larger, the scatter plot develops a longer and longer ``tail'' in OP space, pointing towards the ice-I$c$ region. Hence, small clusters will be rich in ice-0 particles, since ice 0 is the first phase reached by the tail (making justice to the observations of Russo {\it et al.} in Ref.\,\cite{Russo5}); as the cluster size increases, the number of hexagonal-ice particles grows accordingly. At the barrier top, the scatter plot overlaps with cubic-ice and ice-0 regions only marginally, indicating that the number of icelike particles different from hexagonal is percentually small. In particular, the fraction of ice-0 particles will be smaller for LD-C, due to a smaller overlap of the scatter plot with the ice-0 region. More generally speaking, the fractions of the various phases in the nucleus are sensitive to where phase boundaries are exactly placed in OP space, hence the precise fractions should not be trusted overmuch --- only trends are significant. The closeness of ice-0 and liquid clouds in Fig.\,8(b) and (c) is the sign of a structural proximity between ice 0 and liquid, which in turn explains why the preferred location of ice-0 particles in the nucleus is over the surface. The same argument explains why cubic-ice particles are preferentially found in the core of the nucleus: the OP region pertaining to ice I$c$ is the most distant from the liquid region; hence, owing to a low structural matching, no close contact can occur between cubic-ice and liquid particles.

%
%
\begin{figure*}
\begin{center}
\begin{tabular}{ccc}
\includegraphics[width=5cm]{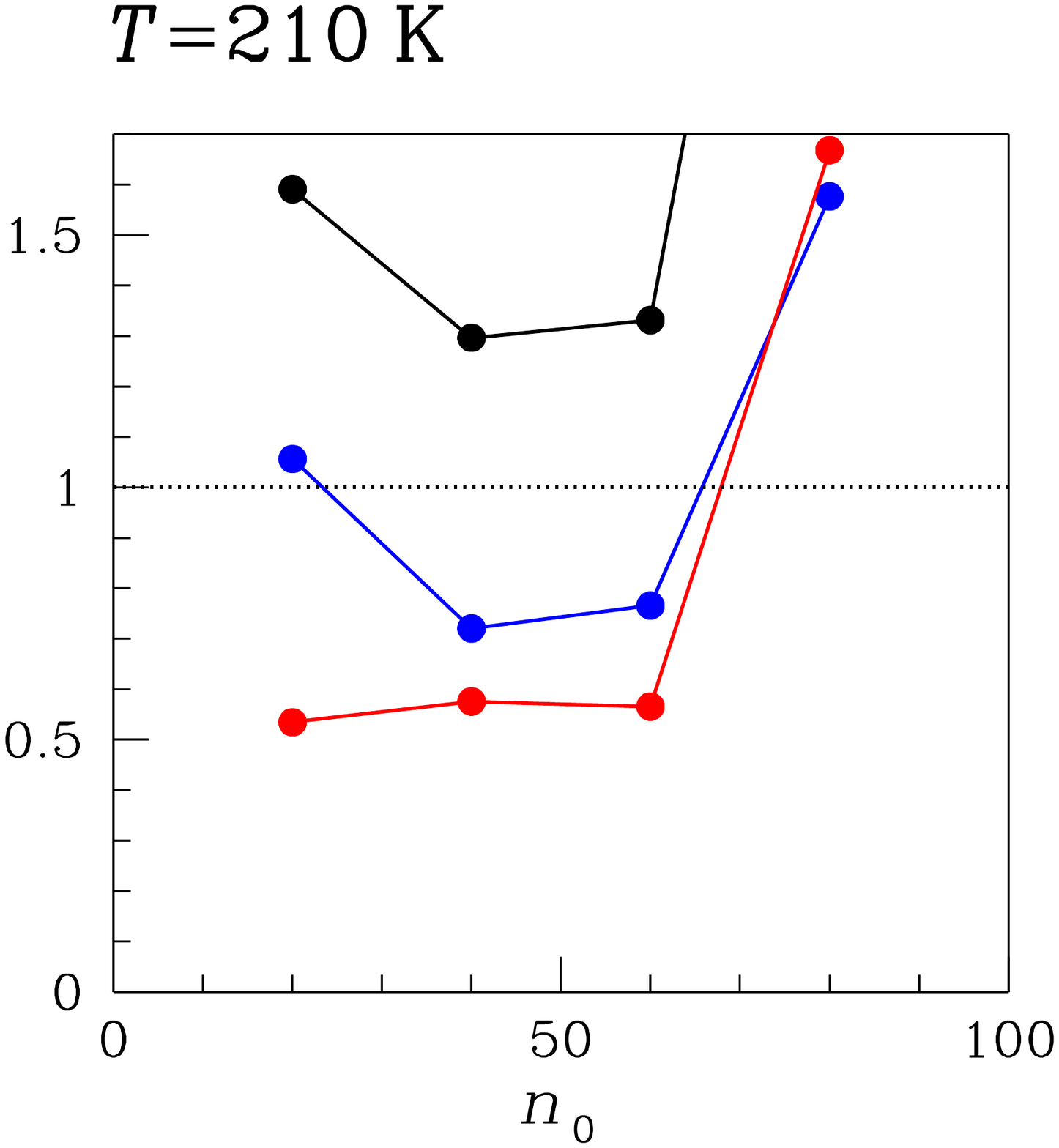} &
\includegraphics[width=5cm]{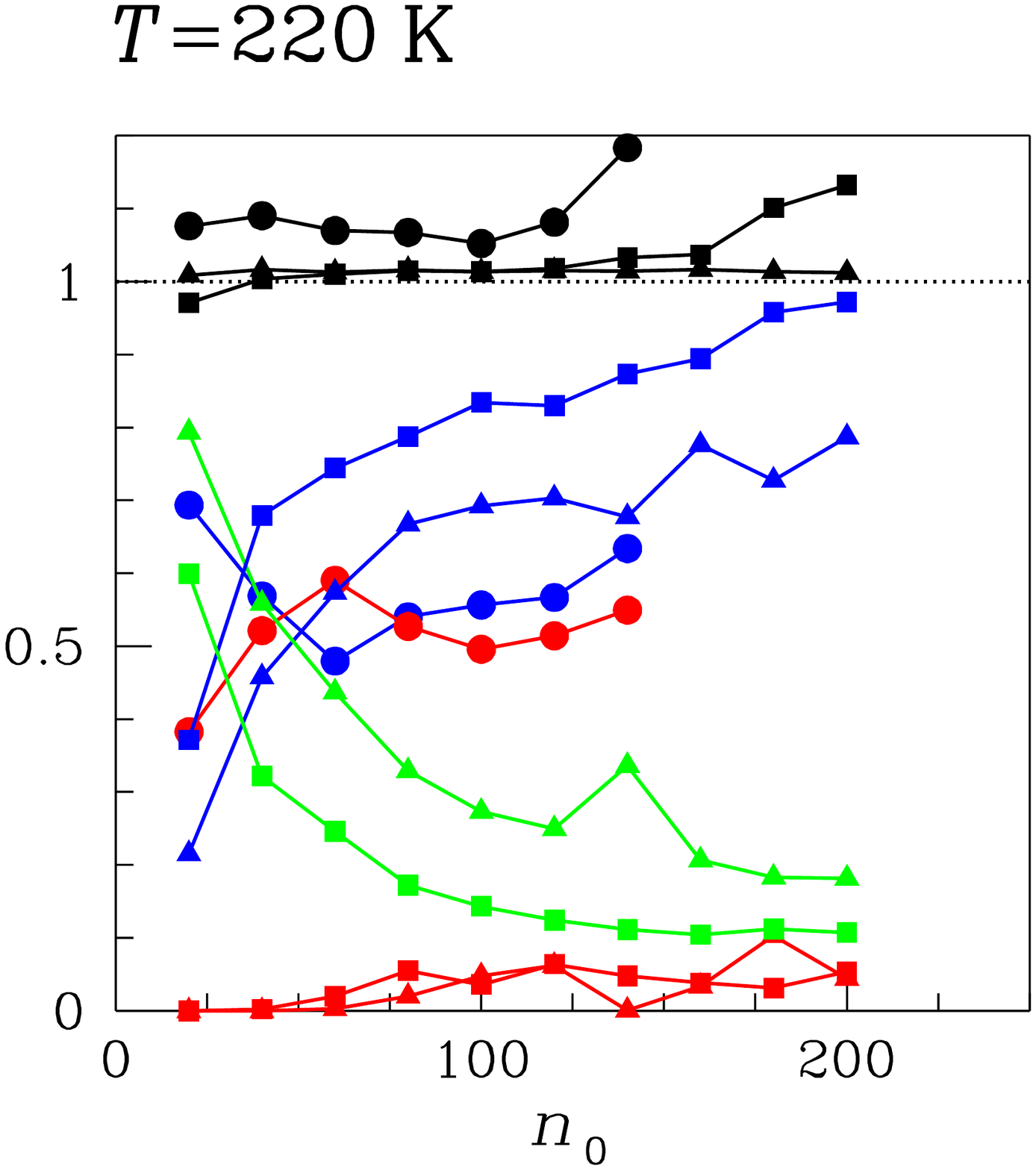} &
\includegraphics[width=5cm]{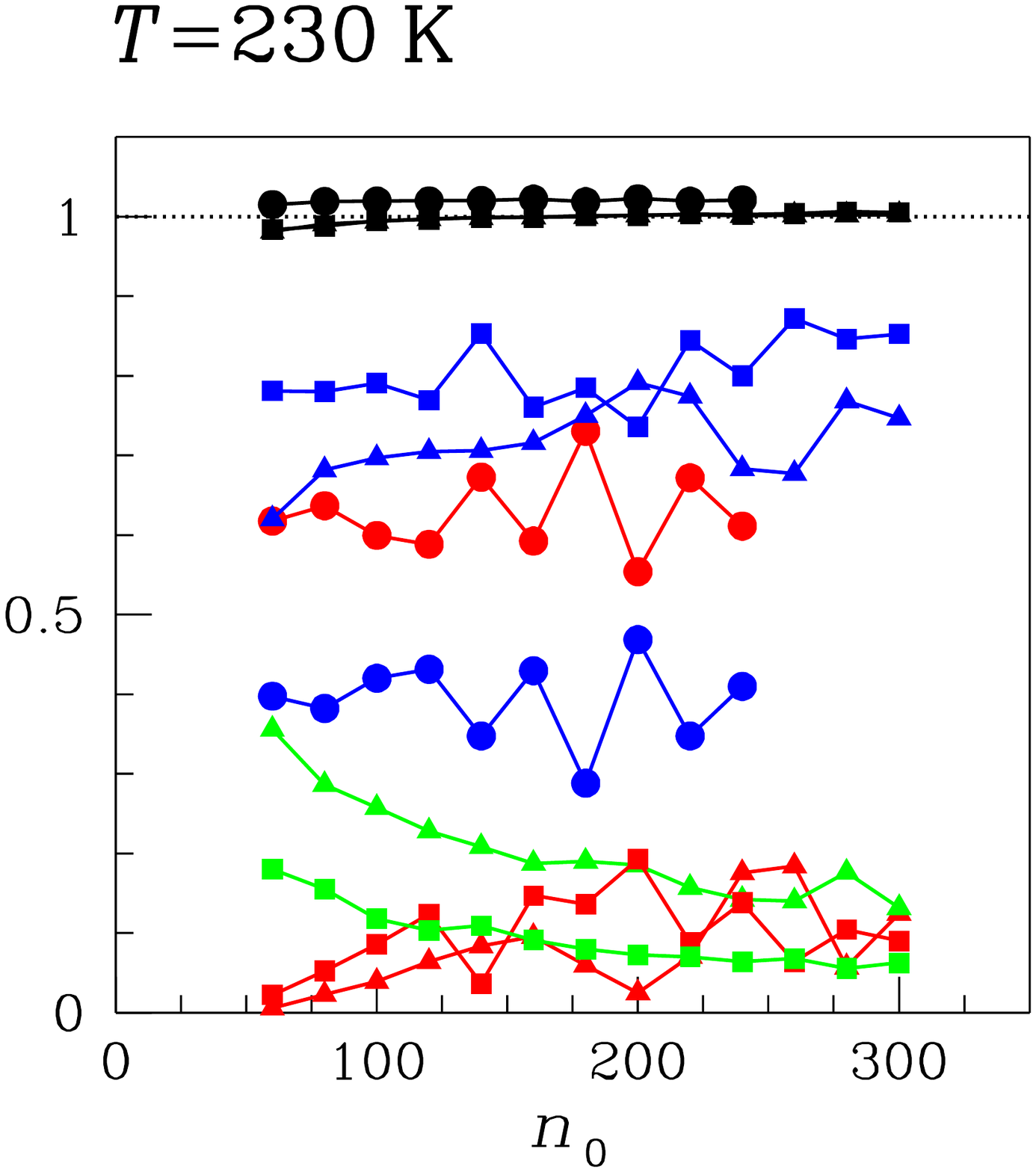}
\end{tabular}
\caption{mW model for $P=0$: percentage of icelike particles of each type (ice I$h$, blue; ice I$c$, red; ice 0, green), for three temperatures (210\,K, left; 220\,K, center; 230\,K, right). Different symbols mark different LD schemes (LD-A, dots; LD-B, triangles; LD-C, squares). More precisely, the plotted data are the ratios of each mean number of particles to $n_0$. Each black datum is obtained by summing the red, blue, and green data. To gather sufficient statistics, I have produced as many as $2\times 10^6$ MC cycles at equilibrium for each $n_0$.}
\label{fig13}
\end{center}
\end{figure*}

Finally, I have monitored during the simulation run the number of icelike particles of each species, wherever they are found in the box, as a function of the size of the maximum cluster. If there were only one cluster in the box, these numbers divided by $n_0$ will sum to one. Starting with LD-A, I first note that for $T=210$\,K the picture of a single cluster does not hold --- see Fig.\,13(a). There is a higher number of hexagonal-ice particles in small clusters, but near criticality icelike particles are almost equally divided between ice I$h$ and ice I$c$. As $T$ increases, cubic-ice particles become eventually prevailing. Moving to the other schemes, Figs.\,13(b) and (c) confirm what has emerged from the analysis of the scatter plot in OP space. For $T=220$\,K, ice-0 particles are the majority in small clusters, but as the cluster gets larger (and more cluster bulk becomes then available) hexagonal order gradually takes over. Cubic-ice particles increase in number with the cluster size, but still remaining a minority. For $T=230$\,K, the fraction of hexagonal-ice particles is already large in small clusters, and keeps large on increasing the size, while the percentage of ice-0 particles decreases. For both 220\,K and 230\,K, the picture of a unique cluster is well observed, at least up to the critical size.

One possibility to make the results for the three LD variables mutually consistent is to admit that the free-energy landscape of mW in a CV space of higher dimensionality has a saddle-point structure comprising nuclei of various nature, very much as observed in Ref.\,\cite{Moroni} for the Lennard-Jones model. In that paper, Moroni {\it et al.} have considered a two-dimensional CV space, finding that crystal nucleation may occur along different pathways, each characterized by a specific kind of nuclei ({\it e.g.}, mostly fcc or of mixed fcc-bcc character). However, a broad saddle point is unlikely to occur for mW, after that Lupi {\it et al.}~\cite{Lupi} have found out a nearly-optimal reaction coordinate (a properly defined cluster size ${\cal N}$, with a committor histogram very close to minimum width) corresponding to nuclei made of randomly stacked cubic and hexagonal layers in a proportion of about 2:1.  At $T=230$\,K the critical size ${\cal N}^*$ is approximately 450, though the size of the nucleus may vary considerably when other suboptimal variables are used instead of ${\cal N}$, at practically no increase in the spread of the committor histogram. This compelling evidence excludes for mW a scenario similar to the LJ fluid, prompting us to look for a different explanation of the present findings.

%
%
\begin{figure}
\includegraphics[width=12cm]{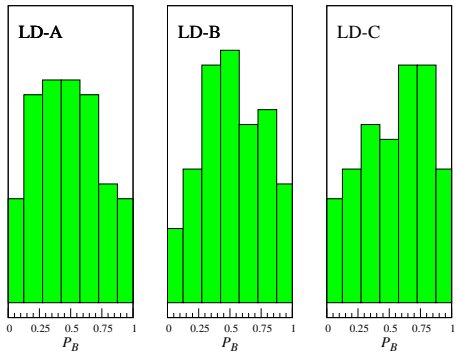}
\caption{mW model for $P=0$: histograms representing the distribution of committor values for the three $n$ variables considered in the present paper. To reduce the noise deriving from a poor statistics, the raw data have been collected in seven bins.}
\label{fig14}
\end{figure}

To clarify things better, for each $n$ variable I have conducted a committor analysis at the top of the barrier, to see whether the alleged critical size is indeed well correlated with the true transition state. The term {\em committor}~\cite{Geissler,Bolhuis} denotes the probability ${\cal P}_B({\bf R}^N)$ that a trajectory initiated at ${\bf R}^N$ will end first in the product phase $B$ (the solid) than in the reactant phase $A$ (the supercooled liquid). The committor provides an exact and universal reaction coordinate, ideal to measure the progress of a transition~\cite{Best,WeinenE}; in practice, nobody knows how to compute the committor analytically for realistic many-particle systems. However, a reasonable test (the histogram test~\cite{Geissler,Pan,Cabriolu}) has been devised, based on the committor concept, to assess {\it a posteriori} the quality of a trial reaction coordinate. In the present case, the tested variable is $n$, the cluster size, as defined through one of three LD prescriptions.

I have first drawn 80 particle configurations of the putative transition-state ensemble from a long MC trajectory running close to $n^*$ (the size of the maximum cluster deviates by at most $\Delta n=4$ from the estimated critical size). Then, 20 independent unconstrained MC trajectories were run from each configuration, recording the fraction $P_B$ of times the system arrives in $B$ before $A$ (the criterion to stop the evolution is a maximum-cluster size smaller or larger than $n^*$ by at least 40). To the extent that $n$ is a good reaction coordinate, the histogram of $P_B$ values will be sharply peaked at 1/2. In Fig.\,14, I show the results obtained for the three $n$ variables. We see that all histograms are very broad, implying that the true transition state encompasses a wide distribution of $n$ values; hence, none of the present LD variables, as reasonable as it may seem, is particularly good as reaction coordinate. Considering that the most likely transition path is not along any of such variables, none can give a fully coherent picture of the nucleation process. Moreover, since the spread of the $P_B$ histogram is similar for all variables, the committor analysis is inconclusive as to what $n$ is best. However, the independent hint from the interface free energy suggests that LD-A actually yields a better reaction coordinate than the other two. Not surprisingly, LD-A also gives a cluster structure more in line with the sophisticated analyses of Refs.\,\cite{Haji-Akbari,Lupi}. In spite of these differences, LD-B and LD-C predict the same barrier height of the LD-A scheme, implying that a certain coordinate $n$ cannot be deemed appropriate for nucleation on the sole basis of the barrier height obtained (though clearly there is no guarantee that the barrier height computed for a non-optimal reaction coordinate like LD-A is really accurate).

%
%
\begin{table}[t]
\caption{Results of a test made to verify whether particle configurations near the top of the barrier for a certain method (first column) are nearly critical also according to another method (second column). For this check, I have employed the same 80 configurations used for the committor analysis ($T=220$\,K). Third column: Average size of the maximum cluster (according to the method in the second column) and standard deviation of the data. Fourth column: mean populations of ices in the maximum cluster. A glance at Fig.\,13 shows that, when the structure of configurations selected near the top of the barrier for a given LD method is analyzed by a different method, not only the critical size but also the ice populations change accordingly.}
\begin{center}
\begin{tabular}{cccc}
\hline\hline
generated by & & \qquad analyzed by & \\
(method, $n$ range) & \qquad (method) & \qquad ($n$ range) & \quad (mean populations) \\
\hline
LD-A, 80-88 & \qquad LD-B & \qquad $132\pm 11$ & \quad 98 (I$h$), 6 (I$c$), 28 (0) \\
LD-A, 80-88 & \qquad LD-C & \qquad $110\pm 10$ & \quad 92 (I$h$), 6 (I$c$), 13 (0) \\
LD-B, 128-136 & \qquad LD-A & \qquad $84\pm 9$ & \quad 55 (I$h$), 29 (I$c$) \\
LD-C, 112-120 & \qquad LD-A & \qquad $88\pm 10$ & \quad 37 (I$h$), 50 (I$c$) \\
\hline\hline
\end{tabular}
\end{center}
\end{table}

To erase any doubt on the correctness of the results obtained by the three LD methods, and rather highlight their actual consistency, I have taken a number of configurations near the top of, say, the LD-A barrier, to see whether they remain close-to-critical when analyzed by the LD-B or LD-C method. The results, collected in Table I, clearly indicate that, despite the bad performance of the three $n$ variables at the histogram test, the nucleus identified with LD-A is one and the same as those identified with LD-B and LD-C (and, conversely, assemblies of configurations that are close-to-critical in terms of LD-B or LD-C retain the same character also according to LD-A). Surprising as it may seem, the difference between these nuclei is then just a matter of particle labeling. This can occur because, as I try to argue in the next paragraph, many particles in the nucleus either have an uncertain status (since they are interfacial particles) or lie near a phase boundary in OP space (see Figs.\,8 and 12). Then, it is not strange if different LD protocols will not agree in the assessment of the nucleus structure.

As a matter of example, consider the configuration depicted in Figs.\,10(a), selected from a LD-A simulation. Its nucleus is a chunk of ice with alternating hexagonal and cubic sheets. When the structure of this configuration is analyzed by the LD-B method, the nucleus changes character, becoming a piece of hexagonal ice with a few ice-0 particles on the surface. On the other hand, when the configuration in Figs.\,10(b) is analyzed by LD-A, the originally hexagonal nucleus is now rich in cubic-like particles, intercalated between hexagonal sheets. This puzzling situation can be explained if we consider the perturbing effect of a heterogeneous and/or disordered particle environment on the values of $\overline{q}_4$ and $\overline{q}_6$ computed by RCUT or NVIC. In other words, if we are not in the bulk, distinct methods of classifying icelike particles may well give different pictures of structurally similar configurations --- especially if the methods are too rough to distinguish {\it e.g.} the cuboctahedron of second neighbors, distinctive of I$c$, from the anticuboctahedron being the footprint of I$h$. In particular, while in LD-A many nucleus particles are interfacial particles living on the boundary between distinct phase regions in OP space, LD-B may simply be unable to discriminate between a purely hexagonal stack and a structure made up of alternating cubic and hexagonal sheets (otherwise, the core of the cluster in Fig.\,10(b) would have been identified as hexagonal also by LD-A). In other words, none of the LD methods considered is sophisticated enough to provide a robust structural identification (then, it does not come as a surprise that the ensuing reaction coordinates are poor).

Summing up, the problem of ice nucleation in mW is paradigmatic of the strong impact that the method of neighbor selection and the related criterion of particle classification can have on the picture of nucleation emerging. It turns out that, while the barrier height is almost insensitive to the protocol used, the critical size and the very same structure of the nucleus are heavily dependent on how solidlike particles are identified and classified. This is especially true for any model of interaction where there is a competition between crystalline phases of nearly equal stability.

\section{Conclusions}

In the present paper, I have investigated for the mW model of water issues related to the nucleation of ice at low pressure. Using Monte Carlo simulation, I have focused on the dependence of cost, size, and structure of the nucleus on the scheme adopted for the classification of particles, for three temperatures near the limit of liquid metastability. I have employed the method put forward by Lechner and Dellago (LD)~\cite{Lechner1}, which, besides distinguishing liquidlike from icelike particles, also allows to identify which type of ice a specific icelike particle belongs to. While the general steps in the application of the LD method are well established, there is still large freedom in the choice of neighboring particles, in turn affecting the values of the OP used to perform the type identification. What clearly emerges for the mW model is that, at strong variance with the LJ model, the details in the implementation of the LD method do matter, proving crucial in shaping the outcome.

I have considered three different LD schemes, denoted A, B, and C. While in LD-A the selected neighbors are the nearest four particles, in the other two schemes the descriptors of orientational order are constructed from particles belonging to the first and second shells (C only differs from B for a more careful choice of phase boundaries in OP space). I find that the cost of nucleus formation is almost the same in all schemes, while the size and structure of the nucleus vary much from one scheme to the other. In LD-A, the nucleus is smaller and its structure prevalently cubic-ice-like. In the other schemes, the critical radius is somewhat larger and the nucleus resembles a piece of hexagonal ice; in fact, there is also a minority of cubic-ice and ice-0 particles, respectively located in the core and on the surface of the nucleus. Nuclei of both types have been repeatedly described in the literature. I argue that the reason for this structural ambiguity is the shortcoming of naive particle-classification methods, which, while good for simple fluids, are inadequate for monatomic water, with the effect of building poor, far-from-optimal nucleation coordinates.

\newpage
{\bf Acknowledgements}
\vspace{5mm}

I would like to thank two anonymous Referees who, by asking the appropriate questions, helped me to considerably improve this study. Simulations were done using the facilities made available at the MIFT department of the University of Messina by the project PO-FESR 2007-2013 MedNETNA (Mediterranean Network for Emerging Nanomaterials).

\appendix

\section{The cluster free energy}
\setcounter{equation}{0}
\renewcommand{\theequation}{A.\arabic{equation}}

CNT assumes the existence of a single solid cluster in the liquid bath, and focuses on the reversible work of formation of this cluster. A unique cluster is a good approximation only near coexistence and when the number of particles is large enough~\cite{tenWolde4}.

Let us first consider the (Landau) free energy of a liquid-plus-solid system of $N$ particles under the constraint that the largest solid cluster has a fixed size $n\ll N$:
\be
F^*_{l+s}(n)\equiv -\frac{1}{\beta}\ln\left\{\int{\rm d}\Gamma\,\delta_{C({\bf R}^N),n}e^{-\beta H(\Gamma)}\right\}\,,
\label{eqA-1}
\ee
where ${\rm d}\Gamma$ is a dimensionless volume element in phase space and $C({\bf R}^N)$ is the size of the largest cluster found in the configuration ${\bf R}^N$ (or the size of the only cluster present if $n\gg 1$). In writing Eq.\,(\ref{eqA-1}), the canonical ensemble is implied, but the nature of the ensemble is actually immaterial for our discussion and any other choice would be equally valid, {\it e.g.}, the isothermal-isobaric ensemble (in this case $H$ is the Hamiltonian plus pressure times volume, and the partition function also includes an integral over volumes). By contrast, the free energy of the pure liquid reads:
\be
F_l=-\frac{1}{\beta}\ln\left\{\int_{C({\bf R}^N)\le n_{\rm max}}{\rm d}\Gamma\,e^{-\beta H(\Gamma)}\right\}\,,
\label{eqA-2}
\ee
under the reasonable assumption that small solid clusters do not alter the prominent liquid nature of the phase ($n_{\rm max}\ll N$). Notice that the integral in Eq.\,(\ref{eqA-2}) is not extended to the whole state space $\Gamma$ but to the domain of liquidlike states only, here identified with those microstates where the size of a solid cluster is at most $n_{\rm max}$, a threshold slightly above the critical size $n^*$. The excess in free energy which results from constraining the size of the maximum cluster is then (for $n\le n_{\rm max}$):
\be
F^*_{\rm cl}(n)=F^*_{l+s}(n)-F_l=-\frac{1}{\beta}\ln\left\{\frac{\int{\rm d}\Gamma\,\delta_{C({\bf R}^N),n}e^{-\beta H(\Gamma)}}{\int_{C({\bf R}^N)\le n_{\rm max}}{\rm d}\Gamma\,e^{-\beta H(\Gamma)}}\right\}=-\frac{1}{\beta}\ln\left\langle\delta_{C({\bf R}^N),n}\right\rangle\,,
\label{eqA-3}
\ee
where the average is taken over liquidlike states only. However, $F^*_{\rm cl}(n)$ is {\em not} the free energy of interest for nucleation theory~\cite{Maibaum}, since the largest cluster is not a generic cluster.

As discussed by many authors in various equivalent ways~\cite{tenWolde2,Bowles,Auer2,Saika-Voivod,Maibaum}, there is a direct correspondence between the free energy cost $F_{\rm cl}(n)$ of a $n$-sized cluster and the cluster-size distribution. Following Maibaum~\cite{Maibaum}, let $c_i({\bf R}^N)$ be the size of the cluster which particle $i$ belongs to, provided $i$ is solidlike, while being $c_i=0$ otherwise. The number $N_n({\bf R}^N)$ of $n$-clusters in the configuration ${\bf R}^N$ is given by
\be
N_n({\bf R}^N)=\frac{\sum_{i=1}^N\delta_{c_i({\bf R}^N),n}}{n}\,\,\,\,\,\,(n\ge 1)\,,
\label{eqA-4}
\ee
and the average number ${\cal N}_n\equiv\left\langle N_n\right\rangle$ of $n$-clusters then reads:
\be
{\cal N}_n=N\frac{\left\langle\delta_{c_1,n}\right\rangle}{n}\,.
\label{eqA-5}
\ee
Denoting $X_s$ the fraction of solidlike particles in the system, the probability that $c_1=n$ given that particle 1 is solidlike equals
\be
P(c_1=n\left|1\,\,{\rm is}\,\,{\rm solidlike}\right.)=\frac{\left\langle\delta_{c_1,n}\right\rangle}{X_s}\,.
\label{eqA-6}
\ee
Then, the cluster free energy (in reduced, $k_BT$ units) is naturally defined to be:
\be
\beta F_{\rm cl}(n)\equiv-\ln\{X_sP(c_1=n\left|1\,\,{\rm is}\,\,{\rm solidlike}\right.)\}+\ln n=-\ln\frac{{\cal N}_n}{N}\,,
\label{eqA-7}
\ee
where proper account has been taken of the translation degeneracy implied by picking up a particular particle in the cluster.

For sufficiently large $n$, all solidlike particles are collected in a single cluster and $c_i({\bf R}^N)=C({\bf R}^N)$ for each particle $i$ in the cluster. Hence, $\sum_{i=1}^N\delta_{c_i({\bf R}^N),n}=C({\bf R}^N)\delta_{C({\bf R}^N),n}=n\delta_{C({\bf R}^N),n}$ and (by Eq.\,(\ref{eqA-4})) $N_n({\bf R}^N)=\delta_{C({\bf R}^N),n}$. As a result,
\be
\beta F_{\rm cl}(n)=\beta F^*_{\rm cl}(n)+\ln N\,.
\label{eqA-8}
\ee
For small $n$ the two sides of Eq.\,(\ref{eqA-8}) yield different numbers and, as argued by Maibaum~\cite{Maibaum}, the latter quantity may be expected to take larger values than the former, due to the penalty involved in constraining the size of the largest cluster to a value smaller than that typical of the metastable liquid.

\section{Umbrella-sampling evaluation of the nucleation barrier}
\setcounter{equation}{0}
\renewcommand{\theequation}{B.\arabic{equation}}

Close to a first-order transition point, no simulation algorithm with local moves is capable to sample the Boltzmann distribution accurately. As a rule, the simulated system gets stuck in the phase-space domain of one phase, being unable to move to the domain of the other phase, which is protected by a high free-energy barrier. The result is an incorrect evaluation of statistical averages (hysteresis). This is exactly the type of difficulty encountered when studying nucleation, since the nucleus falls on top of the saddle between the phases and, as a consequence, the occurrence of a large enough cluster in a moderately supercooled liquid has the character of a rare event.

Umbrella sampling (US) is a method to overcome the limitations of local MC moves~\cite{Torrie1,Torrie2}: it consists in modifying the sampling distribution so as to enhance the occurrence of microstates in the boundary region between the phases, while correcting for the introduced bias in a later moment. Calling $w$ an arbitrary weight function, the average of a statistical estimator $A$ is generally given by:
\be
\left\langle A\right\rangle=\frac{\left\langle A/w\right\rangle_w}{\left\langle 1/w\right\rangle_w}\,,
\label{eqB-1}
\ee
where $\left\langle\cdots\right\rangle_w$ is the average over the Boltzmann distribution multiplied by $w$. In the field of nucleation, a popular choice for $w$ is $\exp(-\beta\widetilde{U})$ with
\be
\widetilde{U}({\bf R}^N)=\frac{1}{2}\kappa\left(C({\bf R}^N)-n_0\right)^2\,,\,\,\,\,\,\,(\kappa>0)
\label{eqB-2}
\ee
which encourages the size of the largest cluster to stay near $n_0$~\cite{tenWolde3}. With this weight, the computation of the Landau free-energy for $C({\bf R}^N)$ turns out to be particularly simple (see Eqs.\,(\ref{eqA-3}) and (\ref{eqB-1})):
\be
\beta F^*_{\rm cl}(n)=-\frac{1}{2}\beta\kappa(n-n_0)^2-\ln\left\langle\delta_{C({\bf R}^N),n}\right\rangle_w+\ln\left\langle\exp\left\{\beta\widetilde{U}({\bf R}^N)\right\}\right\rangle_w\,.
\label{eqB-3}
\ee
In the same way, one immediately obtains from Eq.\,(\ref{eqA-7}):
\be
\beta F_{\rm cl}(n)=\ln N-\ln\left\langle N_n\exp\left\{\beta\widetilde{U}({\bf R}^N)\right\}\right\rangle_w+\ln\left\langle\exp\left\{\beta\widetilde{U}({\bf R}^N)\right\}\right\rangle_w\,.
\label{eqB-4}
\ee
In practice, the constant term on the right-hand side of Eqs.\,(\ref{eqB-3}) and (\ref{eqB-4}) is hard to evaluate since $\exp\{\beta\widetilde{U}\}$ grows fast as one moves farther from $n_0$, whereas sampling is increasingly less accurate. The usual way out is to adopt a divide-and-conquer strategy: 1) A whole sequence of US simulations is carried out, each relative to a different $n_0$ increasing in steps of $\Delta n={\cal O}(1)$ (for each $n_0$, only results in the window $|n-n_0|\le\Delta n$ are taken); 2) successive free-energy branches are matched together continuously; 3) the unknown offset is finally fixed by just observing that the values of $F_{\rm cl}(n)$ and $F^*_{\rm cl}(n)$ for very small $n$ can be accurately obtained in a standard MC simulation.

The rationale behind the above procedure is clear: While the relative frequencies of the various $n$-dependent quantities are estimated correctly near $n_0$, the computed value of $\left\langle\exp\{\beta\widetilde{U}\}\right\rangle_w$ is not reliable. As a result, in the $n$ interval where two adjacent windows overlap the free-energy branches at most differ by a constant. As a practical demonstration, in Fig.\,15 I compare data for the cluster free energy of a Lennard-Jones fluid from three sources: 1) a standard Metropolis MC simulation of the supercooled liquid; 2) US simulations with weight (\ref{eqB-2}); 3) US simulations carried out by using a different $w$, {\it i.e.},
\ba
w({\bf R}^N)=\left\{
\begin{array}{ll}
0 & \quad{\rm for}\,\,\,\left|n-n_0\right|\le\Delta n \\
+\infty & \quad{\rm otherwise.}
\end{array} \right.
\label{eqB-5}
\ea
Note that the US data shown in Fig.\,15 are the raw, unprocessed data, {\it i.e.}, no attempt has been made to match the different branches together and then correct for the offset.

%
%
\begin{figure}
\includegraphics[width=10cm]{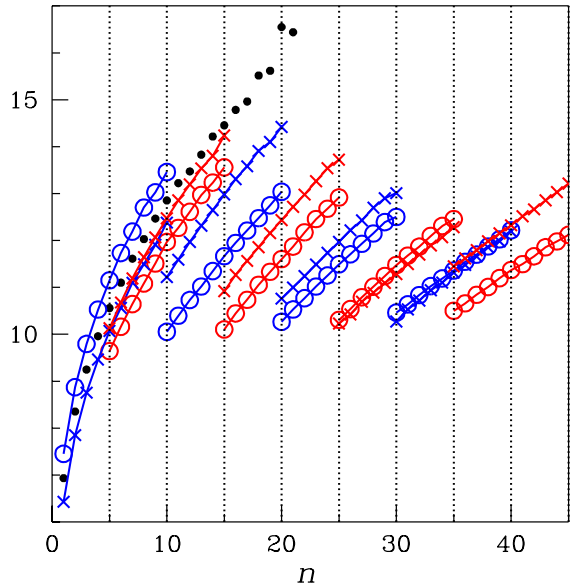}
\caption{In this figure, a comparison is made between three different ways to compute the cluster free energy (in units of $k_BT$) of the cut-and-shifted Lennard-Jones model~\cite{Abramo}, for $T=0.90,P=7.71$, and $N=4000$. The exact $F_{\rm cl}(n)$, obtained from a standard Monte Carlo simulation of the overcompressed liquid (black full dots), is contrasted with the raw data (blue and red) from two sequences of US simulations, which only differ for the choice of the weight function: either $\exp\{-\beta\widetilde{U}\}$, with $\widetilde{U}$ defined at (\ref{eqB-2}) and $\kappa=0.2$ (crosses); or Eq.\,(\ref{eqB-5}) (open dots). In both cases, $n_0=5,10,\ldots,40$ (dotted lines) and $\Delta n=5$. Blue and red colors are alternated only to help the eye.}
\label{fig15}
\end{figure}

A final question to address is how deep into the metastable-liquid region one can go to study nucleation by simulation. Upon cooling the liquid more and more, and provided that equilibrium is attained at each step, the height of the nucleation barrier progressively decreases, implying that the spontaneous appearance of a crystal nucleus becomes increasingly likely. Eventually, one arrives at a temperature where, during a typical run, the nucleation event occurs almost immediately. One has reached the so-called homogeneous nucleation temperature, $T_{\rm H}$, which thus is the first (pressure-dependent) temperature at which crystallization occurs before the liquid equilibrates. Clearly, when a spanning crystal seed appears before equilibrium is established, the very same definition of $F_{\rm cl}(n)$ is lost.

On the other hand, for a given number $N$ of particles in the sample the numerical calculation of $F_{\rm cl}(n)$ cannot be pursued above a certain temperature $T_{\rm max}(N)$, where the diameter of the critical cluster becomes comparable to the side of the simulation box ($T_{\rm max}(N)$ grows with $N$, though remaining smaller than the nominal crystal-liquid transition temperature $T_m$). At this point, the nucleus starts to interact with its periodic images outside the box, if not even becoming a spanning cluster, which in either case brings the numerical procedure to failure. Hence, the computation of $F_{\rm cl}(n)$ can only be carried out in the temperature range between $T_{\rm H}$ and $T_{\rm max}(N)$.

\end{document}